\documentclass[doublespacing]{elsart}

\usepackage{icarus}

\usepackage{natbib}

\bibpunct{(}{)}{;}{a}{,}{,}

\usepackage{graphicx}
\usepackage{lscape}
\usepackage{longtable}
\usepackage[modulo, pagewise]{lineno}

\newcommand{\BibTeX}{ \textrm{B\kern-.05em\textsc{i\kern-.025em b}\kern-.08em
    T\kern-.1667em\lower.7ex\hbox{E}\kern-.125emX} }

\newcounter{ionctr}
\ifx \undefined \ion \newcommand{\ion}[2]{\setcounter{ionctr}{#2}{#1$\;${\small\rmfamily\Roman{ionctr}}\relax}} \fi

\newcommand\arcsec{\mbox{$^{\prime\prime}$}}

\begin{document}

\begin{frontmatter}



\title{Evolution of H$_2$O, CO, and CO$_2$ Production in Comet C/2009 P1 Garradd During the 2011-2012 Apparition\thanksref{McD}}

\thanks[McD]{This paper includes data taken at The McDonald Observatory of The University of Texas at Austin.}

\author[label1]{Adam J. McKay\thanksref{IRTF}},
\author[label1]{Anita L. Cochran},
\author[label2,label3]{Michael A. DiSanti\thanksref{IRTF}},
\author[label2,label4]{Geronimo Villanueva},
\author[label5]{Neil Dello Russo},
\author[label5]{Ronald J. Vervack Jr.},
\author[label6]{Jeffrey P. Morgenthaler},
\author[label7]{Walter M. Harris},
\author[label8]{Nancy J. Chanover}

\address[label1]{University of Texas Austin/McDonald Observatory, 2512 Speedway, Stop C1402 , Austin, TX 78712, (U.S.A);amckay@astro.as.utexas.edu, anita@barolo.as.utexas.edu}
\address[label2]{NASA Goddard Center for Astrobiology, NASA GSFC, Mail Stop 690, Greenbelt, MD 20771 (U.S.A.); Michael.A.Disanti@nasa.gov, Geronimo.Villanueva@nasa.gov}
\address[label3]{Solar System Exploration Division, Mail Stop 690, Greenbelt, MD 20771 (U.S.A)}
\address[label4]{Department of Physics, Catholic University of America, Washington, D.C. 20061 (U.S.A.)}
\address[label5]{Johns Hopkins University Applied Physics Laboratory, 11100 Johns Hopkins Rd., Laurel, MD, 20723 (U.S.A.); neil.dello.russo@jhuapl.edu, Ron.Vervack@jhuapl.edu}
\address[label6]{Planetary Science Institute, 1700 E. Fort Lowell, Ste 106,  Tucson, AZ, 85719 (U.S.A);jpmorgen@psi.edu}
\address[label7]{Lunar and Planetary Laboratory, University of Arizona, 1629 E University Blvd., Tucson, AZ, 85721 (U.S.A);wharris@lpl.arizona.edu}
\address[label8]{Astronomy Department, New Mexico State University, 1320 Frenger Mall, Las Cruces, NM 88001 (U.S.A.); nchanove@nmsu.edu}
\thanks[IRTF]{Visiting Astronomer at the Infrared Telescope Facility, which is operated by the University of Hawaii under Cooperative Agreement no. NNX-08AE38A with the National Aeronautics and Space Administration, Science Mission Directorate, Planetary Astronomy Program.}

\begin{center}
\scriptsize
Copyright \copyright\ 2014 Adam J. McKay, Anita L. Cochran, Michael A. DiSanti, Geronimo Villanueva, Neil Dello Russo, Ronald J. Vervack, Jeffrey P. Morgenthaler, Walter M. Harris, Nancy J. Chanover 
\end{center}


%
%
%
%
%


\end{frontmatter}



\begin{flushleft}
\vspace{1cm}
Number of pages: \pageref{lastpage} \\
Number of tables: \ref{lasttable}\\
Number of figures: \ref{lastfig}\\
\end{flushleft}


\begin{pagetwo}{Evolution of Ices in Comet C/2009 P1 Garradd}

Adam J. McKay \\
University of Texas Austin\\
2512 Speedway, Stop C1402\\
Austin, TX 78712, USA. \\
\\
Email: amckay@astro.as.utexas.edu\\
Phone: (512) 471-1402

\end{pagetwo}


\begin{abstract}
We present analysis of high spectral resolution NIR spectra of CO and H$_2$O in comet C/2009 P1 (Garradd) taken during its 2011-2012 apparition with the CSHELL instrument on NASA's Infrared Telescope Facility (IRTF).  We also present analysis of observations of atomic oxygen in comet Garradd obtained with the ARCES echelle spectrometer mounted on the ARC 3.5-meter telescope at Apache Point Observatory and the Tull Coude spectrograph on the Harlan J. Smith 2.7-meter telescope at McDonald Observatory.  The observations of atomic oxygen serve as a proxy for H$_2$O and CO$_2$.  We confirm the high CO abundance in comet Garradd and the asymmetry in the CO/H$_2$O ratio with respect to perihelion reported by previous studies.  From the oxygen observations, we infer that the CO$_2$/H$_2$O ratio decreased as the comet moved towards the Sun, which is expected based on current sublimation models.  We also infer that the CO$_2$/H$_2$O ratio was higher pre-perihelion than post-perihelion.  We observe evidence for the icy grain source of H$_2$O reported by several studies pre-perihelion, and argue that this source is significantly less abundant post-perihelion.  Since H$_2$O, CO$_2$, and CO are the primary ices in comets, they drive the activity.  We use our measurements of these important volatiles in an attempt to explain the evolution of Garradd's activity over the apparition. 

\end{abstract}



\section{Introduction}

\subsection{Primary Ices in Comets}
\indent Cometary activity is driven by the sublimation of H$_2$O, CO$_2$, and/or CO ice present in the nucleus.  H$_2$O is thought to be the primary driver of activity when comets are closer to the Sun than about 3 AU, though there are exceptions such as 103P/Hartley where CO$_2$ is the main driver~\citep{AHearn2011}.  At larger heliocentric distances, more volatile species (CO$_2$ and/or CO) are the primary drivers, and their sublimation is often invoked to explain distant activity in comets (e.g. C/1995 O1 Hale-Bopp, which exhibited a coma until it reached a heliocentric distance of 28 AU~\citep{Szabo2012}).  However, the transition between H$_2$O and CO$_2$/CO driven activity in comets is poorly understood.\\  

\indent In addition to being the main drivers of cometary activity, H$_2$O, CO$_2$, and CO are typically the most abundant ices present in cometary nuclei.  The relative abundances of these ices in cometary nuclei can reveal details of their formation and evolutionary history.  There is still much debate in the literature whether the abundances of CO and CO$_2$ in comets reflect thermal evolution of cometary nuclei~\citep{BeltonMelosh2009} or whether the observed compositions reflect formation conditions~\citep{AHearn2012}.  The formation of CO$_2$ likely occurs via grain surface interactions of OH and CO, though this reaction is not completely understood~\citep[][and references therein]{AHearn2012}.  Therefore knowledge of the CO and CO$_2$ abundances in comets is paramount for creating a complete picture of cometary composition and differentiating between the effect of formation conditions and subsequent thermal evolution on cometary composition.\\

\indent Both H$_2$O and CO can be observed from the ground in the NIR, while CO is also observable from ground-based sub-mm observations.  Lacking a dipole moment, CO$_2$ has only been observed through its $\nu_3$ vibrational band at 4.26 $\mu$m, which is heavily obscured by the presence of telluric CO$_2$ and therefore cannot be observed from the ground.  This has led to a paucity of observations of this important molecule.  Before 2004, the CO$_2$ abundance had been measured for only a few comets~\citep{Combes1988,Crovisier1997}.  Observations in the past 10 years by space-based platforms such as Spitzer~\citep{ Pittichova2008,Reach2009, Reach2013} and AKARI~\citep{Ootsubo2012}, as well as observations obtained with the Deep Impact spacecraft~\citep{Feaga2007, AHearn2011, Feaga2014}, have resulted in a nearly ten-fold increase in the number of comets with known CO$_2$ abundances and have emphasized the importance of CO$_2$ in comets.  Spitzer is the only one of these IR observatories still in operation, but it is reaching the end of its operational lifetime.  The launch of the James Webb Space Telescope (JWST) in 2018 will reenable observations of CO$_2$ in comets, but not all comets in the inner solar system will be observable due to elongation angle and non-sidereal tracking constraints.  In any case, the limited time available on space-based platforms (as opposed to ground-based telescopes) severely limits the study of CO$_2$ in comets.  Therefore a ground-based proxy for CO$_2$ production in comets is of fundamental importance.\\

\subsection{Atomic Oxygen as a Proxy}
\indent Atomic oxygen is a photodissociation product of H$_2$O, CO$_2$, and CO, and therefore can serve as a viable proxy for these species.  Specifically, observations of the forbidden oxygen lines at 5577, 6300, and 6364~\AA~can reveal the mixing ratios CO$_2$/H$_2$O and CO/H$_2$O in comets.  Past studies have used [\ion{O}{1}]6300 emission to obtain indirect estimates of the H$_2$O production rate for many comets~\citep{Spinrad1982, MageeSauer1990, Schultz1992, Morgenthaler2001, Morgenthaler2007, McKay2012, McKay2014}.  Depending on the wavelength of the dissociating photon, photodissociation of H$_2$O, CO$_2$, and CO can result in the release of an \ion{O}{1} atom in an excited state, either $^1$S or $^1$D.  These excited oxygen atoms then radiatively decay through the 5577~\AA~line ($^1$S) or 6300 and 6364~\AA~lines ($^1$D).\\

\indent The \ion{O}{1} atoms will be preferentially released into the coma in either the $^1$S or $^1$D state depending on the identity of the parent molecule.  Water releases O($^1$S) oxygen at a rate that is 3-8\% of the rate for releasing O($^1$D), whereas for CO$_2$ and CO the rate of O($^1$S) release upon photodissociation is 30-90\% of the O($^1$D) release rate~\citep{Delsemme1980,FestouFeldman1981,BhardwajRaghuram2012}.  These relative efficiencies are reflected in the ratio of the line intensities (hereafter referred to as the ``oxygen line ratio''), given by
\begin{equation}
R\equiv\frac{N(O(^1S))}{N(O(^1D))}=\frac{I_{2972}+I_{5577}}{I_{6300}+I_{6364}}
\end{equation}
where $N(x)$ denotes the column density of the species $x$ and $I_y$ denotes the intensity of line $y$.  In the past calculations of the oxygen line ratio using Eq. 1 have ignored the 2972~\AA~line due to it being much fainter than the other lines (10\% of the 5577~\AA~line~\citep{Slanger2011}) and not being observable from the ground.  As our observations are not sensitive to this line, we will follow this practice when calculating the oxygen line ratios presented in this work.  For sufficiently low number densities where collisional quenching is insignificant, the oxygen line ratio will never be greater than 1, because every atom that decays through the 5577~\AA~line will subsequently decay through the 6300~\AA~or 6364~\AA~line. This is illustrated in Fig.~\ref{energy}, which shows the energy level diagram for \ion{O}{1}.  Therefore a ratio of 0.03-0.08 suggests that H$_2$O is the dominant parent, whereas a ratio of 0.3-0.9 implies that the primary parent molecule is CO$_2$ or CO~\citep{Delsemme1980,FestouFeldman1981,BhardwajRaghuram2012}.  This is a qualitative way of assessing the dominant parent of \ion{O}{1}, and has been employed in the past to show that the dominant parent is H$_2$O~\citep{CochranCochran2001, Cochran2008, Capria2002, Capria2008}.  Recently, it has been suggested that the oxygen line ratio can be used to infer the CO$_2$/H$_2$O ratio in comets, provided that the physics responsible for the release of \ion{O}{1} is understood~\citep{McKay2012, McKay2013, Decock2013}.\\

\indent We present analysis of high resolution NIR and optical spectroscopy of comet C/2009 P1 (Garradd) (hereafter Garradd) obtained during its 2011-2012 apparition.  We employ the NIR spectra to obtain production rates of H$_2$O and CO, and the optical spectra to infer the CO$_2$ and H$_2$O abundance from analysis of the oxygen lines.  The paper is organized as follows.  In section 2 we describe our observations, reduction and analysis procedures.  Section 3 presents our CO, CO$_2$, and H$_2$O production rates and caveats to be considered when interpreting CO$_2$/H$_2$O ratios inferred from the oxygen line ratio.  In section 4 we discuss the implications of our results for the volatile activity of Garradd.  Section 5 presents a summary of our conclusions.

\section{Observations and Data Analysis}
\indent We obtained data on Garradd using three instruments and facilities.  We acquired NIR spectra of Garradd for studying CO and H$_2$O using the CSHELL instrument mounted on the NASA Infrared Telescope Facility (IRTF) on top of Maunakea, Hawaii.  We obtained most of the optical spectra of Garradd for studying atomic oxygen with the ARCES echelle spectrometer mounted on the Astrophysical Research Consortium 3.5-m telescope at Apache Point Observatory (APO) in Sunspot, New Mexico.  We also employed the Tull Coude spectrograph at McDonald Observatory to obtain additional high resolution optical spectra.\\
  
\subsection{CO and H$_2$O -  CSHELL}  
\indent We obtained observations of CO and H$_2$O for Garradd with CSHELL in September-October 2011 and January-March 2012.  CSHELL is a high resolution NIR echelle spectrograph operating at R $\equiv$ $\frac{\lambda}{\Delta\lambda}$ $\sim$ 25,000 with a spectral range of 1-5 $\mu$m.  The detector is a 256 $\times$ 256 pixel InSb CCD, with a spatial pixel scale of 0.2\arcsec/pixel.  CSHELL does not sample the entire available spectral range simultaneously; instead each given setting encompasses only $\sim$ 0.23 percent of the central wavelength.  Thus specific emissions need to be targeted judiciously, and observations of different species are frequently not simultaneous.  However, for our study we used a setting that measured both CO and H$_2$O simultaneously.\\

\indent We provide details for our CSHELL observations of Garradd in Table~\ref{CSHELL}.  We observed a standard star for flux calibration purposes as well as for telluric transmittance correction of the cometary spectra (see below).  The slit length was 30\arcsec, and we oriented the slit east-west for all our observations.  Several slit widths can be employed depending on the desired spectral resolution, with narrower slits providing higher spectral resolution.  We employed the 2\arcsec~wide slit for the comet observations (delivering R $\sim$ 25,000), whereas for the flux standard observations we used a 4\arcsec~wide slit to minimize slit losses of stellar flux (delivering R $\sim$ 13,000).\\

\indent For both stellar and comet observations we employed a standard ABBA observing cadence, with A- and B-beam positions offset by 15\arcsec~along the slit.  We obtained flat fields and dark frames immediately following the final ABBA of each observing sequence (for both star and comet), prior to moving the echelle.  We maintained the comet in the slit using the CCD guider internal to CSHELL. To establish beam positions, we first imaged the comet through the Circular Variable Filter (CVF), which when obtaining spectra transmits only the echelle order closest to blaze. Once we verified that Garradd was centered in the slit for both beam positions (Fig.~\ref{CSHELLslit}), these were marked in the guider field-of-view.  Throughout our spectral observations, we kept the comet at these fiducial positions on the CCD with small manual adjustments to the telescope pointing.  We show a corresponding processed spectral image of Garradd in Fig.~\ref{CSHELLspec}.  Details of the data reduction, including cropping, creation of a bad pixel mask, spatial and spectral rectification of individual frames, and extraction of the spectra, are described elsewhere~\cite[e.g.][and references therein]{DiSanti2014}.\\

\indent We applied a line-by-line radiative transfer model (LBLRTM) for the Earth's atmosphere from~\cite{Clough2005} and~\cite{Villanueva2011} fitted to the observed standard star spectrum to correct for telluric atmospheric absorptions.  We convolved this modeled transmittance function to the resolution of the comet spectra (R $\sim$ 25,000) and scaled it to match the cometary continuum intensity.  Subtracting the scaled transmittance model from the observed comet spectrum isolates molecular emission in excess of the continuum (Fig.~\ref{COfit}, top trace).  For flux calibration, we quantified the spatial profile of the standard star along the slit and obtained a point spread function (PSF).  This allowed us to estimate the slit losses; these were minimal because we used the 4\arcsec~wide slit.\\

\indent We employed a spectral fitting model to extract the fluxes for observed species.  The model employed includes line-by-line g-factors (fluorescence efficiencies) and rotational temperatures for the molecules of interest.  G-factors are calculated using a detailed fluorescence model for each species and referencing a model solar spectrum to account for any Swings Effect present~\citep{Villanueva2011, Villanueva2012b}.  Because the CSHELL spectra do not sample enough lines to measure rotational temperature from the observed spectra, we assume a rotational temperature of 50-60 K for our Garradd observations, based on observations of several molecules in the comet at a similar heliocentric distance, using NIRSPEC at Keck~\citep[see][]{DiSanti2014}.  We show an example fit to the CO setting in Garradd on October 10, 2011, in Fig.~\ref{COfit}.\\

\indent We convert the measured line fluxes to production rates using a Haser Model~\citep{Haser1957}.  For parent species this is given by 
\begin{equation}
n(r)=\frac{Q}{4\pi r^2v}(e^{-\beta r}) 
\end{equation}
Here $n$ is the number density, $r$ is the nucleocentric distance, $Q$ is the production rate, $v$ is the coma expansion velocity, and $\beta$ is the inverse photodissociation scale length, defined as 
\begin{equation}
\beta \equiv \frac{1}{v\tau}
\end{equation}
where $\tau$ is the photodissociation lifetime.  We list the photodissociation lifetimes and g-factors employed in Table~\ref{Haserparent}.  The Haser model is used to obtain the factor $f(x)$~\citep{Yamamoto1981}, which accounts for the number of molecules not included in the slit.  Then the nucleus-centered production rate is given by
\begin{equation}
Q=\frac{F 4\pi\Delta^2}{g\tau f(x)}
\end{equation}
where $Q$ is the production rate, $F$ is the observed flux, $\Delta$ is the geocentric distance of the comet, $g$ is the g-factor, and $\tau$ is the photodissociation timescale.\\

In addition to the aperture correction, a Q-curve analysis is required.  Due primarily to seeing and potential slight drift of the comet over an ABBA sequence, the nucleus-centered production rates always underestimate the total (or global) gas production rate.  To account for this, an analysis technique termed Q-curve analysis~\citep{DelloRusso1998} is employed to calculate the growth factor (GF).  We show an example Q-curve for our Garradd observations in Fig.~\ref{QcurveIR}.  Typical values of GF are a factor of 1.5--2.  For very bright comets, a Q-curve analysis can be done for every species, but for moderately bright comets (like Garradd), only the brightest lines are used in the Q-curve analysis, the results of which are assumed to be applicable to the other species.  It is important to note that for our observations of Garradd, CO and H$_2$O are measured \textit{simultaneously} in the same CSHELL setting (see Fig.~\ref{COfit}).  Therefore our observations provide a robust measure of the abundance ratio CO/H$_2$O that is not dependent on the GF employed, so long as GF is the same for both CO and H$_2$O.\\

\subsection{\ion{O}{1} -  ARCES and Tull Coude Spectrograph}

\indent We obtained most optical spectra of Garradd using the ARCES instrument mounted on the 3.5-meter telescope at APO.  ARCES provides a spectral resolution of R = 31,500 and a spectral range of 3500-10,000~\AA~with no interorder gaps.  This large, uninterrupted spectral range allows for simultaneous observations of all three oxygen lines.  More specifics for this instrument are discussed elsewhere~\citep{Wang2003,McKay2012, McKay2013}.\\

\indent The observation dates and geometries are described in Table~\ref{observations}.  All nights except Feb 27 were photometric, meaning absolute flux calibration of the spectra was possible.  We centered the 3.2\arcsec~$\times$ 1.6\arcsec~slit on the optocenter of the comet.  We used an ephemeris generated from JPL Horizons for non-sidereal tracking of the optocenter.  For short time-scale tracking, the guiding software uses a boresight technique, which utilizes optocenter flux that falls outside the slit to keep the slit on the optocenter.  We observed a G2V star in order to remove the underlying solar continuum and Fraunhofer absorption lines.   We obtained spectra of a fast rotating (vsin(i) $>$ 150 km s$^{-1}$), O, B, or A star to account for telluric features and spectra of a flux standard to establish absolute intensities of cometary emission lines.  The calibration stars used for each observation date are given in Table~\ref{observations}.  We obtained spectra of a quartz lamp for flat fielding and acquired spectra of a ThAr lamp for wavelength calibration.\\

\indent Spectra were extracted and calibrated using IRAF scripts that perform bias subtraction, cosmic ray removal, flat fielding, and wavelength calibration.  We divided each comet, G2V, and flux standard star spectrum by the fast-rotator spectrum to remove telluric features.  We then converted the tellurically corrected comet spectrum flux to physical units using the tellurically corrected flux standard spectrum (for photometric nights).  We assumed an exponential extinction law and extinction coefficients for APO when flux calibrating the cometary spectra~\citep{Hogg2001}.  We shifted the tellurically corrected solar analog spectrum in wavelength to match the comet spectrum.  Then we scaled the solar analog spectrum to the flux calibrated comet spectrum and subtracted the solar analog spectrum to remove absorption lines from the solar continuum reflected off of dust particles.\\

\indent Because of the small size of the ARCES slit, it is necessary to obtain an estimate of the slit losses to achieve an accurate flux calibration.  We find the transmittance through the slit by performing aperture photometry on the slit viewer images as described in~\cite{McKay2014}.  The transmittance is typically between 70-95\% and the typical standard deviation in the transmittance estimate is approximately 10\%.  Therefore we adopt a 10\% systematic uncertainty in our absolute flux calibration.\\

\indent The Tull Coude spectrograph is mounted on the 2.7-meter Harlan J. Smith Telescope at McDonald Observatory.  It provides a spectral resolution of R=60,000 and a spectral range of 3500-10000~\AA.  Although there are interorder gaps redward of 5800~\AA, we took care to set the grating so that the red oxygen lines were encompassed by our observations.  The Tull Coude observations and subsequent data reduction are very similar to those for ARCES.  The one exception is that the Tull Coude spectrograph has a solar port that feeds reflected sunlight from the daytime sky directly into the spectrograph, thereby providing an observed solar spectrum for removal of solar absorption lines and the continuum from the cometary spectra.  More details on reduction of Tull Coude data can be found in~\cite{CochranCochran2001}.

\indent For both ARCES and Tull Coude observations, the atomic oxygen lines are also present as telluric emission features, so a combination of high spectral resolution and large geocentric velocity (and therefore large Doppler shift) is needed to resolve the cometary line from the telluric feature.  For the observations reported here, only on UT August 28 are the telluric and cometary features not sufficiently separated.  However, at this time Garradd was bright enough so that the 6300~\AA~line was much stronger than the telluric feature, so we can use the measured 6300~\AA~line flux to estimate the H$_2$O production rate, with the caveat that there is likely a small ($<$ 10\%) correction needed to account for telluric contamination of the line flux.  However, this assumption is not valid for the 5577~\AA~line, so we do not report an oxygen line ratio on this date.  For the observations where the telluric and cometary [\ion{O}{1}] emission were sufficiently separated, we deblended the lines using the Gaussian-fitting method described in~\cite{McKay2012,McKay2013}.  We show an example spectrum of the 5577~\AA~line in Garradd on September 21 in Fig.~\ref{OIspec}.  The flux ratio of the 6300 and 6364~\AA~lines is well established by both theory and observation to be 3.0~\citep{SharpeeSlanger2006, CochranCochran2001, Cochran2008, McKay2012, McKay2013, Decock2013}, and we confirmed that our derived flux ratio for the 6300 and 6364~\AA~lines was consistent with this value before conducting further analysis.\\ 

\indent To determine H$_2$O production rates from our [\ion{O}{1}]6300~\AA~line observations, we created a simple model of the
expected radial distribution of [\ion{O}{1}]6300~\AA~from all expected sources of this line.  We employed algorithms based on those used in~\cite{Morgenthaler2001,Morgenthaler2007} and~\cite{McKay2012,McKay2014}, which are described in detail in the aforementioned references and are summarized as follows.  We calculate the number density for the species of interest as a function of nucleocentric distance using the computationally simple Haser Model~\citep{Haser1957}.  We modify the Haser scale lengths following the prescriptions of~\cite{Combi2004} to emulate the more physical vectorial model~\citep{Festou1981}, which accounts for isotropic ejection of daughter species following dissociation of the parent molecule.  We derive the expansion velocity of the coma using the~\cite{Tseng2007} relation of gas expansion velocity versus heliocentric distance, which for our observations results in assumed expansion velocities of 0.6-0.8 km s$^{-1}$.  It is important to note that because of our small projected slit size, a large fraction of the gas may not be accelerated to the terminal value calculated using the~\cite{Tseng2007} relation, so our H$_2$O production rates may be slightly biased by this effect.  The physical parameters we employ for each molecule are given in Table~\ref{Haser}.   The contribution of CO$_2$ to the [\ion{O}{1}]6300~\AA~flux is provided by our (model dependent) oxygen line ratio calculations (see below).  We find that the derived H$_2$O production rates are not particularly sensitive to the assumed CO$_2$ abundance, and any small changes in H$_2$O production rates from assuming different values of the CO$_2$ production rate are well within our uncertainties in flux calibration.  All photodissociative lifetimes are adopted from~\cite{Huebner1992} and are given for a heliocentric distance of 1 AU.\\

\indent The oxygen line ratio is related to the ratios of the column densities of major oxygen-containing species.  Following~\cite{McKay2012},  
\begin{equation}
\frac{N_{CO_2}}{N_{H_2O}} = \frac{RW^{red}_{H_2O}-W^{green}_{H_2O}-W^{green}_{CO}\frac{N_{CO}}{N_{H_2O}}+RW^{red}_{CO}\frac{N_{CO}}{N_{H_2O}}}{W^{green}_{CO_2}-RW^{red}_{CO_2}}
\end{equation}
where $N$ is column density and $R$ is the oxygen line ratio.  The release rate $W$ is defined as $W \equiv \tau^{-1}\alpha\beta$, where $\tau$ represents the photodissociative lifetime of the parent molecule, $\alpha$ is the yield into the excited state of interest, and $\beta$ represents the branching ratio for a given line out of a certain excited state.  If the contribution of CO photodissociation to the \ion{O}{1} population (in both $^1$D and $^1$S states) is considered negligible~\citep{RaghuramBhardwaj2014}, Eq. 5 simplifies to~\citep{McKay2013}:
\begin{equation}
\frac{N_{CO_2}}{N_{H_2O}} = \frac{RW^{red}_{H_2O}-W^{green}_{H_2O}}{W^{green}_{CO_2}-RW^{red}_{CO_2}}
\end{equation}
For a FOV much smaller than the photodissociation scale length of the parent species (this applies to both ARCES and Tull Coude observations), the production rate $Q$ is given by
\begin{equation}
Q=<N>vd 
\end{equation}
where $<N>$ is the average column density in the FOV in molecules/cm$^2$, $v$ is the expansion velocity of the gas, and $d$ is the radius of the observing aperture.  Since $v$ and $d$ are the same for the two species (de facto for $d$, an assumption for $v$), the production rate is directly proportional to the column density, so the ratio of column densities in the slit FOV is also the ratio of production rates.  Because the lifetime $\tau$ depends on heliocentric distance, in principal the values of $W$ depend on heliocentric distance.  However, assuming the H$_2$O, CO$_2$, and CO lifetimes all scale the same way (i.e. $r^{-2}$), this dependence cancels out in Eqs. 5 and 6, so any results derived from Eqs. 5 and 6 are independent of the scaling of the values of $\tau$.\\

\indent The utility of Eqs. 5 and 6 is limited by the accuracy to which the release rates $W$ are known.  Unfortunately, laboratory data for the $W$ values is lacking~\citep{Huestis2008}.  Values given in the literature (mostly theoretical in nature) vary by a factor of 2-3.  Therefore employing Eqs. 5 and 6 results in systematic uncertainties in the derived value of CO$_2$/H$_2$O in addition to the stochastic uncertainties from the measurement.  This needs to be kept in mind when interpreting the oxygen line ratio in terms of a quantitative measure of the CO$_2$/H$_2$O ratio.\\

\indent There are several assumptions needed for Eqs. 5 and 6 to be valid.  First, photodissociation of H$_2$O, CO$_2$, and CO must be the only sources of $^1$S and $^1$D \ion{O}{1} atoms.  This is usually the case, as shown by~\cite{FestouFeldman1981}.  The more uncertain assumption is that radiative decay is the only loss mechanism for $^1$S and $^1$D \ion{O}{1} atoms.  These states may also be de-excited via collisions with H$_2$O.  At number densities typical of cometary comae and FOV associated with ground-based observations, collisional de-excitation (quenching) will not serve as a significant sink for $^1$S \ion{O}{1} atoms.  However, collisional quenching can be a significant loss mechanism for $^1$D \ion{O}{1} atoms, especially in the innermost coma and for large production rates (Q$_{H_2O}$ $>$ 10$^{30}$ mol s$^{-1}$)~\citep{RaghuramBhardwaj2014}.  Because the projected slit size for our observations is $\sim$ 1000 km, we are sampling the inner coma and therefore collisional quenching of $^1$D \ion{O}{1} atoms could be significant.\\

\indent Therefore we performed additional analysis to account for preferential collisional quenching of $^1$D atoms as compared to $^1$S atoms.  The oxygen line ratio employed in Eqs. 5 and 6 assumes the ratio was calculated using 6300~\AA~and 6364~\AA~line intensities that are unaffected by collisional quenching.  Since this may not be the case, the observed 6300~\AA~and 6364~\AA~line intensities need to be increased to account for the $^1$D atoms that were de-excited through collisions and thus do not contribute to the 6300~\AA~and 6364~\AA~line intensities.  We estimate the percentage of atoms lost to collisional quenching using the Haser Model for $^1$D \ion{O}{1} described above, which includes collisional quenching of $^1$D \ion{O}{1} atoms~\citep{Morgenthaler2001,McKay2012,McKay2014}.  We first calculate the H$_2$O production rate using the observed 6300~\AA~flux with collisional quenching turned on.  We then run another Haser Model with this production rate with collisional quenching turned off.  The difference between the predicted flux from the model without collisional quenching and the observed flux then gives an estimate of how much collisional quenching is present.  This factor is then used to scale up the observed 6300~\AA~and 6364~\AA~line intensities when calculating the oxygen line ratio.  We determined that for our Garradd observations this scale factor was dependent on both geocentric distance and H$_2$O production rate, and found values ranging from 1.1-1.5, with the largest values corresponding to smaller geocentric distances and large production rates.  This effect dominates our stochastic error for most of our observations (at large heliocentric distance, the stochastic errors and collisional effects are comparable); therefore not accounting for collisional quenching can add systematic error to the inferred CO$_2$/H$_2$O ratios.\\

\subsection{Uncertainties}
\indent We note that all uncertainties quoted in this work include 1-sigma stochastic errors, which for these observations are dominated by Poisson statistics of the cometary spectra.  For absolute production rates, the uncertainties also include systematic error associated with flux calibration (as discussed above), which is the dominate source of error for H$_2$O production rates derived from \ion{O}{1} emission.  We have not included systematic error associated with the uncertainty of the \ion{O}{1} release rates adopted in the formal error bars, but discuss the effect of this on our results at length in Section 4.  

\section{Results}
\indent In this section we present the CO and H$_2$O production rates (or upper limits) measured from our CSHELL observations.  We also present H$_2$O production rates derived from our [\ion{O}{1}]6300~\AA~observations and inferred CO$_2$/H$_2$O ratios derived from the oxygen line ratio.  As discussed in Section 2.2, the release rates needed to infer the CO$_2$ abundance from \ion{O}{1} observations are not known to an accuracy of better than a factor of three.  In this section we present the motivation for the particular release rates we adopt and we discuss the consequences of adopting different release rates in the next section.\\

\indent As described in Section 2.1, we employed our CSHELL infrared observations of H$_2$O and CO to provide a direct measurement of the production rates of these species.  The results, with their 1-sigma error bars, are summarized in Table~\ref{CSHELLQrates}.  The CO/H$_2$O mixing ratios derived from these measurements are also shown in Table~\ref{CSHELLQrates} and plotted as a function of heliocentric distance as circles in Figure~\ref{COmixrat}.  In March we did not detect H$_2$O with CSHELL, therefore our H$_2$O production rate for March is a 3-sigma upper limit and correspondingly CO/H$_2$O is a 3-sigma lower limit.  In Fig.~\ref{COmixrat} we show a calculated CO/H$_2$O ratio assuming the H$_2$O production rate inferred from \ion{O}{1} emission.  It is clear that the CO/H$_2$O ratio is much higher post-perihelion than pre-perihelion, a result also observed by others~\citep{Feaga2014,Bodewits2014}.\\

\indent The resulting H$_2$O production rates for the ARCES observations are given in Table~\ref{AveQrates} and plotted with the CSHELL H$_2$O production rates as a function of heliocentric distance in Fig.~\ref{QH2O_opt}.  The pre-perihelion H$_2$O production rates derived from the CSHELL observations are higher than those from ARCES, while those from post-perihelion observations are inconclusive.  We present a possible explanation for this discrepancy in section 4.3.  The measured oxygen line ratios are given in Table~\ref{OIratio}, both measured and corrected for collisional quenching.  The uncertainty is particularly small for the R=2 AU post-perihelion data point because a large number of observations of the \ion{O}{1} line ratio were made on this date, driving down the stochastic error (this applies to the inferred CO$_2$/H$_2$O ratio as well).  The values corrected for collisional quenching are plotted versus heliocentric distance in Fig.~\ref{OIratiofig}.\\

\indent As noted in Section 2.2, the CO$_2$/H$_2$O ratio inferred from the oxygen line ratio is dependent on the adopted release rates $W$.  Fortunately, the CO$_2$/H$_2$O ratio was measured directly by the EPOXI spacecraft very close in time (mere days) from our March observations~\citep{Feaga2014}.  The EPOXI results show CO$_2$/H$_2$O $\sim$ 8 $\pm$ 2\%.  The \ion{O}{1} release rates from~\cite{BhardwajRaghuram2012} are likely the most accurate to date available in the literature.  However, applying these release rates in Eq. 5 gives CO$_2$/H$_2$O $\sim$ 2\%, much less than observed by~\cite{Feaga2014}.  Given that \ion{O}{1} release rates remain poorly constrained, for the rest of this paper we will adopt empirical values that we have found are able to reproduce the CO$_2$/H$_2$O ratio in Garradd measured by~\cite{Feaga2014}.  We give these release rates and those from~\cite{BhardwajRaghuram2012} in Table~\ref{Wrates} and the inferred CO$_2$/H$_2$O ratios using our adopted release rates in Table~\ref{OIratio}.  We plot the CO$_2$/H$_2$O ratios in Fig.~\ref{COmixrat}.  Preliminary analysis suggests that the adopted \ion{O}{1} release rates reproduce the CO$_2$ abundance in comet 103P/Hartley 2 observed by the EPOXI mission~\citep{AHearn2011} as well, and in a future publication we will examine in more detail how well these empirical release rates reproduce the observed CO$_2$ abundances for 103P/Hartley and other comets.  However, for the present work we note that these release rates are consistent with the CO$_2$ abundance measured in Garradd at a heliocentric distance of 2 AU post-perihelion~\citep{Feaga2014}, and thus we employ them here for our observations.  In the next section we will discuss in more detail the effect of adopting different release rates on our results.  It is important to note that the release rates we employ for the rest of this paper are strictly empirical (and may not represent a unique solution), i.e. they reproduce current observations, but there is no physical mechanism known to explain why they should be different from those of~\cite{BhardwajRaghuram2012}.  More work is certainly needed to establish robust release rates.\\

\section{Discussion}
In this section, we discuss the behavior of the CO/H$_2$O (section 4.1) and CO$_2$/H$_2$O (section 4.2) ratios over Garradd's apparition.  In section 4.3 we discuss the discrepency between H$_2$O production rates measured by CSHELL and ARCES, and section 4.4 will discuss the implications of our results for a possible picture of Garradd's primary ices.\\

\subsection{Asymmetry in CO/H$_2$O With Respect to Perihelion}
\indent One key finding of this work is the strong asymmetry in the CO/H$_2$O ratio with respect to perihelion in comet Garradd.  Other observers have also seen this phenomenon~\citep{Feaga2014, Bodewits2014}.  Feaga et al. (2014) measured a very high value for the CO/H$_2$O ratio of 63\%, when the comet was at R=2.0 AU post-perihelion.  Our non-detection of H$_2$O with CSHELL at this time places a lower limit of 18.2\% on the CO/H$_2$O ratio.  In Section 3 we derived a more meaningful CO/H$_2$O value in late March of $\sim$ 40\% by using the H$_2$O production rate inferred from our \ion{O}{1} measurements.  Despite the potential uncertainties in the derivation of our water production rate in late March, the primary reason for the difference between our CO/H$_2$O value and that of~\cite{Feaga2014} is likely due to the difference in treatment of optical depth effects in the CO measurements.~~\cite{Feaga2014} accounted for optical depth affects using the radiative transfer model of~\cite{GerschAHearn2014}, whereas we ignored optical depth effects.~\cite{GerschAHearn2014} found that for Garradd and our projected slit size at the comet ($\sim$ 1000 km), optical depth could decrease the effective g-factor by more than 30\%.  This means that if the optically thin g-factors are employed (as we have done), our CO production rate will be an underestimate.  Reducing the g-factor we employ by 30\% results in much better agreement between~\cite{Feaga2014} and this work.  In any case, our observations support the finding from~\cite{Feaga2014} that the CO/H$_2$O ratio was much higher post-perihelion than pre-perihelion.\\

\indent The increase of the CO/H$_2$O ratio is a manifestation of H$_2$O and CO production having very different behaviors with respect to perihelion: H$_2$O production is higher pre-perihelion while CO production increases throughout the apparition, even as the comet is receding from the Sun.  This is the first time this peculiar evolution of CO production has been observed in a comet.  One explanation for this is that most of the CO in Garradd is buried at depth, and it is only post-perihelion that the thermal wave propagates far enough into the nucleus to fully activate CO~\citep{Bodewits2014}.  Another possible explanation is that there is a seasonal effect, and one of Garradd's rotational poles is more abundant in CO than the other one.  When the CO-rich pole receives more direct insolation, it becomes fully activated and the CO production rate increases.  This theory has some validity because Garradd has a large obliquity of about 60$^{\circ}$~\citep{Farnham2013}.\\

\indent Finally, it is interesting to note that despite the large change in the CO/H$_2$O ratio over the apparition, the oxygen line ratio is much more constant over the time period covered by our CSHELL observations (though it does change slightly, see the next section).  This suggests that the oxygen line ratio is not very sensitive to the CO/H$_2$O abundance, as is expected based on the current understanding of CO photochemistry and recent modelling efforts~\citep{RaghuramBhardwaj2014}.  However,~\cite{RaghuramBhardwaj2014} did find that for large CO abundances comparable to the H$_2$O abundance, the contribution of CO could have some effect on the oxygen line ratio.  This work and~\cite{Feaga2014} find a CO/H$_2$O abundance of 40-60\% at a heliocentric distance of 2 AU post-perihelion, suggesting that a CO contribution could have a measureable effect on the oxygen line ratio at this time.  This implies that the CO abundance we measure for Garradd could affect our inferred CO$_2$ abundance (see Eq. 5).  We will discuss this further in the following section.\\

\subsection{CO$_2$/H$_2$O}
\indent Our inferred CO$_2$/H$_2$O ratio decreases as Garradd moves toward perihleion.  This is expected based on the relative volatilities of H$_2$O and CO$_2$~\citep{MeechSvoren2004}.  This trend is also observed in the oxygen line ratios measured by~\cite{Decock2013} for Garradd, which we compare to our measured oxygen line ratios in Fig~\ref{OIratiofig}.  We applied a collisional quenching correction to the~\cite{Decock2013} values using our methodology.  Since~\cite{Decock2013} do not report H$_2$O production rates, we assumed our derived values for the most contemporaneous observations.\\

\indent The values measured by~\cite{Decock2013} are systematically lower than our values, even after collisional quenching has been accounted for (see Fig.~\ref{OIratiofig}).  One possibility is that because the UVES slit employed by~\cite{Decock2013} is much narrower than the ARCES and Tull Coude slits, the icy grain source of H$_2$O for Garradd (see next section and~\cite{Combi2013,DiSanti2014,Bodewits2014}) may have contributed less to the H$_2$O production in the UVES slit.  Therefore by employing our H$_2$O production rates to calculate the collisional quenching correction we may be overcorrecting the measured oxygen line ratio measured by~\cite{Decock2013} when comparing to our measurements (i.e. a lower H$_2$O production rate should have been employed).  More detailed knowledge of the possible icy grain source (size, spatial distribution, etc.) would be needed to confirm or refute this possibility.  Another possible explanation is that~\cite{Decock2013} attempted to subtract C$_2$ contamination from the 5577~\AA~line, while we have assumed that the contribution from C$_2$ is negligible.  This would result in their oxygen line ratios being systematically lower than ours.  For their observations,~\cite{Decock2013} applied a small ($\sim$ 10\%) correction for C$_2$ contamination for their observation at a heliocentric distance of 2 AU, but did not observe any potentially contaminating C$_2$ emission at larger heliocentric distances, meaning they did not apply any correction for C$_2$ emission for the observations at larger heliocentric distances (Alice Decock, private communication).  Therefore C$_2$ contamination may account for the small discrepancy near 2 AU, but cannot account for the potential discrepency at 2.5 AU.  As the C$_2$ contamination is small to non-existent for this comet, we have chosen to ignore the possible contribution of C$_2$ when inferring CO$_2$ abundances from our observations because the systematic error associated with the \ion{O}{1} release rates is much larger than that associated with C$_2$ contamination in this case, and so will not affect our conclusions.\\

\indent When deriving our CO$_2$ abundances via Eq. 5, we employed our measured CO/H$_2$O ratios when possible (i.e. dates for which we have near contemporaneous observations of CO with CSHELL), and applied CO/H$_2$O values from~\cite{Yang2012} and~\cite{Feaga2014} for larger heliocentric distances for which we obtained only optical data.  However, we noted in the previous section that there is a discrepancy between our measured value of CO/H$_2$O in March as compared to that measured by~\cite{Feaga2014}. This is likely due to radiative transfer effects, which we considered negligible in our analysis.  For low CO abundances ($<$ 20\%), adopting different CO values has negligible effects on the derived CO$_2$ abundances, but for the large values of CO/H$_2$O observed post-perihelion for Garradd, such effects could be substantial.  If we employ our adopted release rates from Table~\ref{Wrates} and the CO/H$_2$O ratio of 63 $\pm$ 21\% found by~\cite{Feaga2014} in Eq. 5, we derive a CO$_2$ abundance of 5.7 $\pm$ 1.9\% (error bars are 1-sigma and stochastic), as compared to 8.5 $\pm$ 2.1\% (also 1-sigma error bar) measured by~\cite{Feaga2014}.  There is significant overlap of the 1-sigma error bars, but it is enough of a discrepancy to warrant further investigation.\\

\indent To bring the inferred CO$_2$/H$_2$O ratio in line with that measured by~\cite{Feaga2014} using their measured CO abundance requires lowering our adopted CO$_2$ release rates in the last column of Table~\ref{Wrates} by a factor of 1.5, while keeping the H$_2$O and CO release rates unchanged (in principal one could change any of the different release rates; we chose CO$_2$ because release rates for this molecule are the least constrained).  This raises the CO$_2$/H$_2$O ratio to 8.7 $\pm$ 2.9\%, clearly consistent with the CO$_2$/H$_2$O ratio measured by~\cite{Feaga2014}.  In Table~\ref{CO2comparison} we show how the modified CO$_2$ release rates change its inferred abundance for our other dates of observation, and also the CO$_2$ abundance assuming the release rates of~\cite{BhardwajRaghuram2012}.  We find that the overall effect of the modified CO$_2$ release rates is higher CO$_2$ abundances, whereas the~\cite{BhardwajRaghuram2012} release rates give lower CO$_2$ abundances.  However, the~\cite{BhardwajRaghuram2012} release rates do not reproduce the~\cite{Feaga2014} observations, as mentioned previously in Section 3.\\

\indent Accounting for possible systematic errors, the CO$_2$/H$_2$O ratios inferred from our observations seem to be 50-100\% higher pre-perihelion than post-perihelion.  Although it is true that the inferred CO$_2$/H$_2$O ratios are sensitive to the release rates adopted, any variability inferred in this ratio, at least in a qualitative sense, is independent of the adopted release rates.  Indeed, in the specific case examined above, the effect of adopting a higher CO abundance and lower CO$_2$ release rates (or any other release rates that reproduce the~\cite{Feaga2014} CO$_2$ abundances) is to make the pre-/post-perihelion asymmetry in CO$_2$/H$_2$O more pronounced.  Employing the release rates from~\cite{BhardwajRaghuram2012} results in the same qualitative conclusions concerning a decreasing CO$_2$/H$_2$O ratio on the pre-perihelion leg of the orbit and also the asymmetry in the CO$_2$/H$_2$O ratio with respect to perihelion.  Therefore, while characterization of the asymmetry in the CO$_2$/H$_2$O ratio in a quantitative sense is dependent on the exact \ion{O}{1} release rates adopted, the qualitative finding that the CO$_2$ abundance is higher pre-perihelion than post-perihelion is not dependent on the adopted release rates.\\

\indent The asymmetry in the CO$_2$/H$_2$O ratio is the opposite of that observed in the CO abundance; CO/H$_2$O is higher post-perihelion but CO$_2$/H$_2$O is higher pre-perihelion.  If the CO contribution is assumed to be completely negligible in Eq. 5, even for very high CO abundances, then the asymmetry in CO$_2$/H$_2$O with respect to perihelion would disappear.  However, a completely negligible contribution from CO, while probable for low CO abundances, is not likely for high CO abundances~\citep{RaghuramBhardwaj2014}.  Measurements of the relevant release rates in the laboratory will remove this uncertainty, and when those become available the validity of the inferred asymmetry in the CO$_2$/H$_2$O abundance with respect to perihelion can be reexamined.\\

\subsection{Aperture Effect for H$_2$O Production Rates}
\indent In Section 3 we noted that the H$_2$O production rates for our CSHELL observations were systematically higher than those derived from the ARCES observations, at least pre-perihelion.  For pre-perihelion observations, there is a notable discrepancy between H$_2$O production rates derived from IR slit spectroscopy~\citep{Paganini2012, Villanueva2012,DiSanti2014} vs. wide-field imaging in OH~\citep{Bodewits2014} and Lyman-$\alpha$~\citep{Combi2013}.  This has been attributed to an extended source of H$_2$O that was incompletely sampled by IR slit spectroscopy but was completely sampled by the wide field imaging~\citep{Combi2013,DiSanti2014,Bodewits2014}.  The leading candidate for this source is icy grains.  In addition to the aperture effect,~\cite{DiSanti2014} determined that the pre-perihelion H$_2$O spatial profile was highly asymmetric, which they interpret as being caused by an extended source of H$_2$O in the projected sunward-facing hemisphere, likely due to sublimation from icy grains.\\

\indent Both our ARCES and CSHELL observations of H$_2$O support the hypothesis that icy grains were an important source of H$_2$O production for Garradd, at least pre-perihelion.  We plot the derived H$_2$O production rate at a heliocentric distance of 2 AU pre-perihelion as a function of aperture size in Fig.~\ref{aperture}.  Our derived H$_2$O production rates from the ARCES observations are consistent with those determined via IR slit spectroscopy with NIRSPEC and CRIRES~\citep{Paganini2012, Villanueva2012,DiSanti2014}.  However, our CSHELL-derived H$_2$O production rates are systematically higher than both our H$_2$O production rates from the [\ion{O}{1}]6300~\AA~line flux and other IR measurements, but not as high as the production rates from wide field imaging of OH and Lyman-$\alpha$.  We believe this is because the CSHELL slit employed was 2\arcsec~ wide, while the slit widths in the other IR investigations with NIRSPEC and CRIRES were much narrower ($<$ 0.5\arcsec~).  By virtue of its larger slit width (compared to ARCES or other IR observations) and longer slit (compared to ARCES), CSHELL sampled a larger fraction of the coma than either the ARCES or other IR observations.  Therefore we would expect a larger derived H$_2$O production rate from the CSHELL observations if an extended source of H$_2$O is important.  At the same time, the CSHELL H$_2$O production rates should be less than those derived from wide-field imaging, which is what is observed.\\

\indent We also present in Fig.~\ref{aperture} the H$_2$O production rates as a function of aperture size when Garradd was at a heliocentric distance of 2 AU post-perihelion.  The aperture effect noted pre-perihelion seems to still be present, but at a smaller level.  Only the~\cite{Combi2013} H$_2$O production rate at an aperture size of $\sim$ 10$^7$ km shows a large deviation, and the observations from~\cite{Feaga2014} and~\cite{Bodewits2014} with aperture sizes of approximately 10$^5$ km are only slightly higher than our ARCES derived value.  The upper limit on the H$_2$O production rate from our CSHELL observations is not particularly constraining, but is consistent with the other data points.  The decrease in the aperture effect suggests that post-perihelion H$_2$O production from icy grains was less important than it was pre-perihelion.  This is also consistent with the nearly symmetric spatial profile and lower production rate measured for H$_2$O at 1.57 AU post-perihelion~\citep{DiSanti2014}.\\

\subsection{A Possible Picture for Garradd's Primary Ices}
\indent The evolution of the production rates of H$_2$O, CO$_2$, and CO over the course of the apparition possibly reveals much about the evolution of Garradd's activity, and the volatiles responsible for that activity.  For R=2-3 AU pre-perihelion, the observed CO/H$_2$O ratio was typically 5\%~\citep{Bodewits2014}, while the CO$_2$/H$_2$O ratio dropped from 35\% to 12\% (see Table~\ref{OIratio}) over this range.  Therefore it seems likely that CO$_2$ was the driver of activity on the inbound leg, and our ARCES observations followed the transition from CO$_2$- to H$_2$O-driven activity.  Post-perihelion, the CO/H$_2$O ratio was much larger, while the CO$_2$/H$_2$O ratio was smaller, suggesting that CO had a much larger role in driving the activity post-perihelion.\\

\indent The asymmetry in the CO$_2$/H$_2$O ratio may explain the evolution of the icy grain H$_2$O source observed both by our observations and those of others.  For comet 103P/Hartley,~\cite{AHearn2011} found that much of the H$_2$O production came from icy grains that were accelerated into the coma by outgassing CO$_2$.  If this is a general phenomenon in cometary sublimation and this effect is present for Garradd, then the reduction in the extended source of H$_2$O post-perihelion can be described in terms of a reduction in CO$_2$ production.  Less CO$_2$ outgassing means less icy grains, which translates into the observed reduction in H$_2$O production.\\

\indent It is less clear what caused the reduction in the CO$_2$ release to begin with.  One possibility is depletion of CO$_2$ in the outgassing layers, but it seems that CO would also be depleted by this same mechanism, contrary to the much higher CO production rates observed post-perihelion.  Another possibility is large scale compositional heterogeneity in Garradd's nucleus.  An active area rich in CO$_2$ was preferentially exposed to sunlight pre-perihelion and controlled the sublimation behavior at that time.  Post-perihelion, another region rich in CO was exposed and drove the activity at that time.  As mentioned earlier,~\cite{Farnham2013} found a high obliquity for Garradd's nucleus, making a seasonal interpretation of changes in mixing ratios in terms of pole orientation more plausible.~~\cite{Bodewits2014} found that the H$_2$O production from the nucleus (i.e. sublimating directly from the nucleus and not from icy grains in the coma), was symmetric with respect to perihelion, meaning that the nucleus-production of H$_2$O was not sensitive to the changing illumination conditions caused by the high obliquity of the nucleus.\\

\indent If correct, the seasonal effect in CO and CO$_2$ caused by compositional heterogeneity and a high obliquity has several implications for Garradd's activity and (perhaps) cometary composition in general.  If CO$_2$ is responsible for releasing icy grains into the coma, then CO$_2$, as observed with 103P/Hartley, is an important driver of cometary activity, even inside the H$_2$O ice line.  CO may also play a role if it is abundant enough relative to H$_2$O and CO$_2$, as seen with Garradd post-perihelion.  As a significant amount of the H$_2$O production is outgassing directly from the nucleus, H$_2$O still plays a significant role in driving cometary activity inside of 3 AU from the Sun, but H$_2$O may not dominate the activity as is typically assumed, with CO$_2$ and (to a lesser extent) CO playing an important role.  This makes cometary activity a complicated phenomenon, with multiple sublimating ices likely driving activity at all points in the orbit.\\

\indent If Garradd does exhibit large-scale heterogeneity in the CO$_2$ and CO content of its ices, this could imply that Garradd consists of two or more cometisimals that formed in disparate regions of the protosolar disk.  One part of the nucleus might contain CO-rich cometisimals that formed out near the CO ice line, whereas other parts might consist of cometisimals that formed closer to the CO$_2$ ice line and are comparitively rich in CO$_2$ and less abundant in CO.  This is consistent with the early Solar System being a turbulent place, with mixing of material from various regions of the protosolar disk contributing to the formation of planetary bodies.\\

\section{Conclusions}
We present analysis of observations of H$_2$O (directly, and indirectly via \ion{O}{1} emission), CO (directly), and CO$_2$ (indirectly via \ion{O}{1} emission) in comet C/2009 P1 Garradd throughout its 2011-2012 apparition.  We observed an asymmetry in the CO/H$_2$O ratio with respect to perihelion, a result observed by others~\citep{Feaga2014,Bodewits2014}.  We observe that the oxygen line ratio (and therefore the CO$_2$/H$_2$O ratio) decreased as the comet approached perihelion, which was also observed by~\cite{Decock2013}.  We also observe an asymmetry in the CO$_2$/H$_2$O ratio with respect to perihelion, though since these determinations are from indirect observations of CO$_2$ using \ion{O}{1} emission, this result is sensitive to our understanding of the photochemistry responsible for the release of \ion{O}{1} into the coma.  The observed asymmetry in the CO$_2$/H$_2$O ratio is different from that observed for CO/H$_2$O: CO$_2$/H$_2$O is higher pre-perihelion, but CO/H$_2$O is higher post-perihelion.  This may suggest that there is large scale heterogeneity in Garradd's nucleus, with different ices driving the activity at different points in the orbit.  The observed variability in the coma composition of Garradd highlights the need to observe individual comets throughout their entire apparitions.  Long-term observing campaigns can reveal insights into cometary composition that are missed by single (snap-shot) observations.  Our analysis demonstrates the power of employing observations of \ion{O}{1} in comets to study primary ice abundances (namely CO$_2$) and sublimation activity.  Laboratory measurements and additional contemporaneous observations of H$_2$O, CO, CO$_2$, and \ion{O}{1} in comets are necessary to constrain the photochemistry of \ion{O}{1} release from H$_2$O, CO$_2$, and CO.\\

\ack
We thank the two anonymous reviewers whose comments improved the quality of this manuscript.  This work was supported by the NASA GSRP Fellowship program through grant number NNX11AO03H and by the NASA Planetary Atmospheres Program through grant number NNX08A052G.  We thank John Barentine, Jurek Krzesinski, Chris Churchill, Pey Lian Lim, Paul Strycker, and Doug Hoffman for developing and optimizing the ARCES IRAF reduction script used to reduce these data.  We acknowledge the NASA-Infrared Telescope Facility for their support of our Garradd CSHELL observations and the APO observing specialists for their assistance with the Garradd ARCES observations.  We would also like to acknowledge the JPL Horizons System, which was used to generate ephemerides for nonsidereal tracking of the comets during the observations, and the SIMBAD database, which was used for selection of reference stars.  The authors wish to recognize and acknowledge the very significant cultural role and reverence that the summit of Maunakea has always had within the indigenous Hawaiian community.  We are most fortunate to have the opportunity to conduct observations from this mountain.

\label{lastpage}


\bibliography{../references.bib}

\bibliographystyle{plainnat}

\clearpage

\begin{table}
\caption{Log of CSHELL Observations
\label{CSHELL}
}
\begin{tabular}{ccccc}
Date (UT) & $r$ (AU) & $\Delta$ (AU)& $\dot{\Delta}$ (km s$^{-1}$) & Standard Star\\
\hline
September 19-21, 2011 & 2.01 & 1.57 & 18.6 & HR 6556\\
October 10-12, 2011 & 1.85 & 1.81 & 19.4 & HR 6556\\
January 26-27, 2012 & 1.62 & 1.63 & -23.2 & HR 6324\\ 
February 27-28, 2012 & 1.79 & 1.28 & -7.6 & HR 5054\\
March 21-22, 2012 & 1.96 & 1.36 & 20.0 & HR 3888\\
March 28, 2012 & 2.01 & 1.44 & 25.7 & HR 3888\\
\hline
\end{tabular}
\end{table}

\begin{table}
\caption{Adopted Scale Lengths and G-factors for CSHELL Analysis
\label{Haserparent}
}
\begin{tabular}{lcccc}
Molecule & $\tau_p$ (s)$^a$ & g-factor (ergs s$^{-1}$ molecule$^{-1}$)\\
\hline
H$_2$O & 8.3 $\times$ 10$^4$ & 2.6 $\times$ 10$^{-13}$\\
CO & 1.3 $\times$ 10$^6$ & 2.3 $\times$ 10$^{-14}$\\
\hline
\end{tabular}\\
$a$ adopted from~\cite{Huebner1992} and given for $R$=1 AU\\
$b$ adopted from~\cite{Villanueva2012b} for H$_2$O and from~\cite{DiSanti2003} for CO, given for $R$=1 AU\\
\end{table}

\clearpage

\begin{table}
\caption{Log of ARCES and Tull Coude Observations
\label{observations}
\label{lasttable}
}
\begin{tabular}{llllllllll}
\hline
Date (UT) & $r$ (AU) & $\Delta$ (AU) & $\dot{\Delta}$ (km s$^{-1}$) & G2V & Fast Rot. & Flux Cal\\
\hline
June 18-19, 2011 & 2.88 & 2.50 & -45.9 & G 93-22 & HR 8826 & HR 8634\\
July 30, 2011 & 2.47 & 1.57 & -25.6 & HD 182081 & HR 8231 & BD +28 4211\\
August 27, 2011 & 2.20 & 1.40 & -9.7 & HD 177082 & 7 Vul & 58 Aql\\
September 14-15, 2011* & 2.06 & 1.51 & 16.4 & solar port & HR 8419 & -\\
September 21, 2011 & 2.00 & 1.58 & 18.8 & HD 177082 & 7 Vul & 58 Aql\\
October 10-12, 2011 & 1.85 & 1.81 & 19.4 & HD 177082 & 7 Vul & 58 Aql\\
November 4, 2011 & 1.69 & 2.03 & 11.9 & G 93-22 & 7 Vul & 58 Aql\\
February 3, 2012* & 1.65 & 1.52 & -22.6 & solar port & Alpha Lyrae & -\\
February 27, 2012 & 1.79 & 1.28 & -8.1 & HD 129920 & HR 5693 & -\\
March 28-29, 2012 & 2.02 & 1.46 & 27.2 & LTT 12303 & HR 3958 & HD 93521\\
April 28, 2012 & 2.28 & 2.11 & 43.2 & HD 76617 & HR 3586 & HD 93521\\
\hline
\end{tabular}
*obtained with Tull Coude spectrograph at McDonald Observatory
\end{table}

\clearpage

\begin{table}
\begin{center}
\caption{Parameter Values Used in the Haser Models for \ion{O}{1}
\label{Haser}
}
\begin{tabular}{lccc}
Molecule & $\alpha$ & $\tau$ (s)$^a$ & $V_{ej}$ (km s$^{-1}$)$^{b}$\\
\hline
H$_2$O$^c$ & 0.05 & 8.3 $\times$ 10$^4$ & -\\
H$_2$O$^d$ & 0.855 & 8.3 $\times$ 10$^4$ & -\\
~~~~OH & 0.094 & 1.3 $\times$ 10$^5$ & 0.98\\
CO$_2$ & 0.72 & 5.0 $\times$ 10$^5$ & -\\
\hline
\end{tabular}
\end{center}
$a$ Given for $r$=1 AU\\
$b$ Only applicable for \ion{O}{1} that comes from OH photodissociation.  Value from~\cite{Crovisier1989} and~\cite{WuChen1993}.\\
$c$ For dissociation of H$_2$O into H$_2$ and O\\
$d$ For dissociation of H$_2$O into H and OH 
\end{table}

\clearpage
\begin{table}
\begin{center}
\caption{H$_2$O and CO Production Rates from CSHELL Observations
\label{CSHELLQrates}
\label{lasttable}
}
\begin{tabular}{lllll}
\hline
 & & \multicolumn{2}{c}{$Q$ (10$^{28}$ mol s$^{-1}$)} &\\
UT Date & $R$ (AU) & CO & H$_2$O & CO/H$_2$O ($\%$)\\
\hline
9/21/2011 & 2.01 & 0.67 $\pm$ 0.09 & 14.1 $\pm$ 3.8 & 4.6 $\pm$ 1.1\\
10/10/2011 & 1.85 & 1.03 $\pm$ 0.18 & 16.4 $\pm$ 3.9 & 6.2 $\pm$ 1.1\\
1/25/2012 & 1.62 & 1.64 $\pm$ 0.17 & 10.5 $\pm$ 2.2 & 14.7 $\pm$ 3.2\\
2/27/2012 & 1.69 & 1.96 $\pm$ 0.33 & 10.0 $\pm$ 3.1 & 19.5 $\pm$ 5.0\\
3/28/2012 & 2.01 & 1.55 $\pm$ 0.14 & $<$ 8.5$^a$ & $>$ 18.2$^b$\\
\hline\\
\end{tabular}
\end{center}
$a$ 3-$\sigma$ upper limit\\
$b$ 3-$\sigma$ lower limit, if Q$_{H_2O}$ from the ARCES observations is adopted, then the value is 40.8 $\pm$ 5.5$\%$\\
\end{table}

\clearpage
\begin{table}
\begin{center}
\caption{H$_2$O Production Rates from ARCES Observations
\label{AveQrates}
\label{lasttable}
}
\begin{tabular}{lll}
\hline
UT Date & $R$ (AU) & Q$_{H_2O}$ (10$^{28}$ mol s$^{-1}$)\\
\hline
6/18/2011 & 2.88 & 0.80 $\pm$ 0.08\\
7/30/2011 & 2.47 & 2.28 $\pm$ 0.23\\
8/28/2011 & 2.20 & 5.60 $\pm$ 0.56\\
9/20/2011 & 2.00 & 7.50 $\pm$ 0.75\\
10/10/2011 & 1.85 & 9.67 $\pm$ 0.97\\
11/4/2011 & 1.69 & 10.6 $\pm$ 1.06\\
3/28/2012 & 2.02 & 3.80 $\pm$ 0.38\\
4/28/2012 & 2.28 & 3.28 $\pm$ 0.33\\
\hline
\end{tabular} 
\end{center}
\end{table}

\clearpage
\begin{table}
\begin{center}
\caption{Oxygen Line Ratios and Inferred CO$_2$/H$_2$O Ratios
\label{OIratio}
\label{lasttable}
}
\begin{tabular}{lllll}
\hline
UT Date & $R$ (AU) & \ion{O}{1} line ratio$^a$ & \ion{O}{1} line ratio$^b$ & CO$_2$/H$_2$O$^c$\\
\hline
6/18/2011 & 2.88 & 0.160 $\pm$ 0.022 & 0.148 $\pm$ 0.02 & 0.347 $\pm$ 0.052\\
7/30/2011 & 2.47 & 0.130 $\pm$ 0.015 & 0.112 $\pm$ 0.013 & 0.238 $\pm$ 0.031\\
9/14/2011 & 2.02 & 0.081 $\pm$ 0.003 & 0.062 $\pm$ 0.002 & 0.112 $\pm$ 0.004\\
9/20/2011 & 2.00 & 0.086 $\pm$ 0.006 & 0.064 $\pm$ 0.004 & 0.116 $\pm$ 0.009\\
10/10/2011 & 1.85 & 0.088 $\pm$ 0.005 & 0.065 $\pm$ 0.004 & 0.117 $\pm$ 0.009\\
11/4/2011 & 1.69 & 0.067 $\pm$ 0.003 & 0.051 $\pm$ 0.002 & 0.086 $\pm$ 0.004\\
2/3/2012 & 1.65 & 0.079 $\pm$ 0.005 & 0.059 $\pm$ 0.004 & 0.097 $\pm$ 0.009\\
2/27/2012 & 1.79 & 0.063 $\pm$ 0.006 & 0.042 $\pm$ 0.004 & 0.056 $\pm$ 0.009\\
3/28/2012 & 2.02 & 0.073 $\pm$ 0.003 & 0.059 $\pm$ 0.002 & 0.080 $\pm$ 0.006\\
4/28/2012 & 2.28 & 0.082 $\pm$ 0.006 & 0.071 $\pm$ 0.005 & 0.108 $\pm$ 0.012\\
\hline
\end{tabular}
\end{center}
$a$ measured value\\
$b$ corrected for collisional quenching\\
$c$ Inferred using Eq. 5, our adopted release rates $W$ from Table~\ref{Wrates}, and the oxygen line ratios corrected for collisional quenching.
\end{table}

\clearpage
\begin{table}
\begin{center}
\caption{Adopted \ion{O}{1} Release Rates
\label{Wrates}
\label{lasttable}
}
\begin{tabular}{llll}
\hline
Parent & \ion{O}{1} State & $W^a$ & $W^b$\\
\hline
H$_2$O & $^1$S & 2.6 & 0.64\\
H$_2$O & $^1$D & 84.4 & 84.4\\
CO$_2$ & $^1$S & 72.0 & 50.0\\
CO$_2$ & $^1$D & 120.0 & 75.0\\
CO & $^1$S & 4.0 & 4.0\\
CO & $^1$D & 5.1 & 5.1\\
\hline
\end{tabular} 
\end{center}
$a$ Release rates in 10$^{-8}$ s$^{-1}$ from~\cite{BhardwajRaghuram2012}\\
$b$ Our adopted empirical release rates in 10$^{-8}$ s$^{-1}$. 
\end{table}

\clearpage

\clearpage
\begin{table}
\begin{center}
\caption{Inferred CO$_2$ Abundances for Several Sets of Release Rates
\label{CO2comparison}
\label{lasttable}
}
\begin{tabular}{lllll}
\hline
UT Date & $R$ (AU) & CO$_2$/H$_2$O$^a$ & CO$_2$/H$_2$O$^b$ & CO$_2$/H$_2$O$^c$\\
\hline
6/18/2011 & 2.88 & 0.347 $\pm$ 0.052 & 0.526 $\pm$ 0.079 & 0.214 $\pm$ 0.038\\
7/30/2011 & 2.47 & 0.238 $\pm$ 0.031 & 0.361 $\pm$ 0.046 & 0.136 $\pm$ 0.022\\
9/14/2011 & 2.02 & 0.112 $\pm$ 0.004 & 0.169 $\pm$ 0.007 & 0.048 $\pm$ 0.003\\
9/20/2011 & 2.00 & 0.116 $\pm$ 0.009 & 0.176 $\pm$ 0.013 & 0.052 $\pm$ 0.006\\
10/10/2011 & 1.85 & 0.117 $\pm$ 0.009 & 0.178 $\pm$ 0.013 & 0.052 $\pm$ 0.006\\
11/4/2011 & 1.69 & 0.086 $\pm$ 0.004 & 0.130 $\pm$ 0.006 & 0.031 $\pm$ 0.003\\
2/3/2012 & 1.65 & 0.097 $\pm$ 0.0095 & 0.146 $\pm$ 0.013 & 0.038 $\pm$ 0.006\\
2/27/2012 & 1.79 & 0.056 $\pm$ 0.009 & 0.084 $\pm$ 0.014 & 0.010 $\pm$ 0.006\\
3/28/2012 & 2.02 & 0.080 $\pm$ 0.006 & 0.087 $\pm$ 0.029 & 0.023 $\pm$ 0.005\\
4/28/2012 & 2.28 & 0.108 $\pm$ 0.012 & 0.129 $\pm$ 0.033 & 0.045 $\pm$ 0.008\\
\hline
\end{tabular}
\end{center}
$a$ Inferred using Eq. 5, our adopted release rates $W$ from Table~\ref{Wrates}, and our CSHELL CO abundances (or adopted from~\cite{Yang2012} and~\cite{Feaga2014})\\
$b$ Inferred using Eq. 5, our adopted release rates $W$ from Table~\ref{Wrates} with the value for the CO$_2$ release rates decreased by a factor of 1.5, and the CO abundance from~\cite{Feaga2014} for March and April.\\
$c$ Inferred using Eq. 5, the release rates from~\cite{BhardwajRaghuram2012}, and our CSHELL CO abundances (or adopted from~\cite{Yang2012} and~\cite{Feaga2014})\\
\end{table}

\clearpage

\begin{center}
 Figure Captions
\end{center}

\begin{figure}[p!]
\begin{center}
\includegraphics[width=\linewidth]{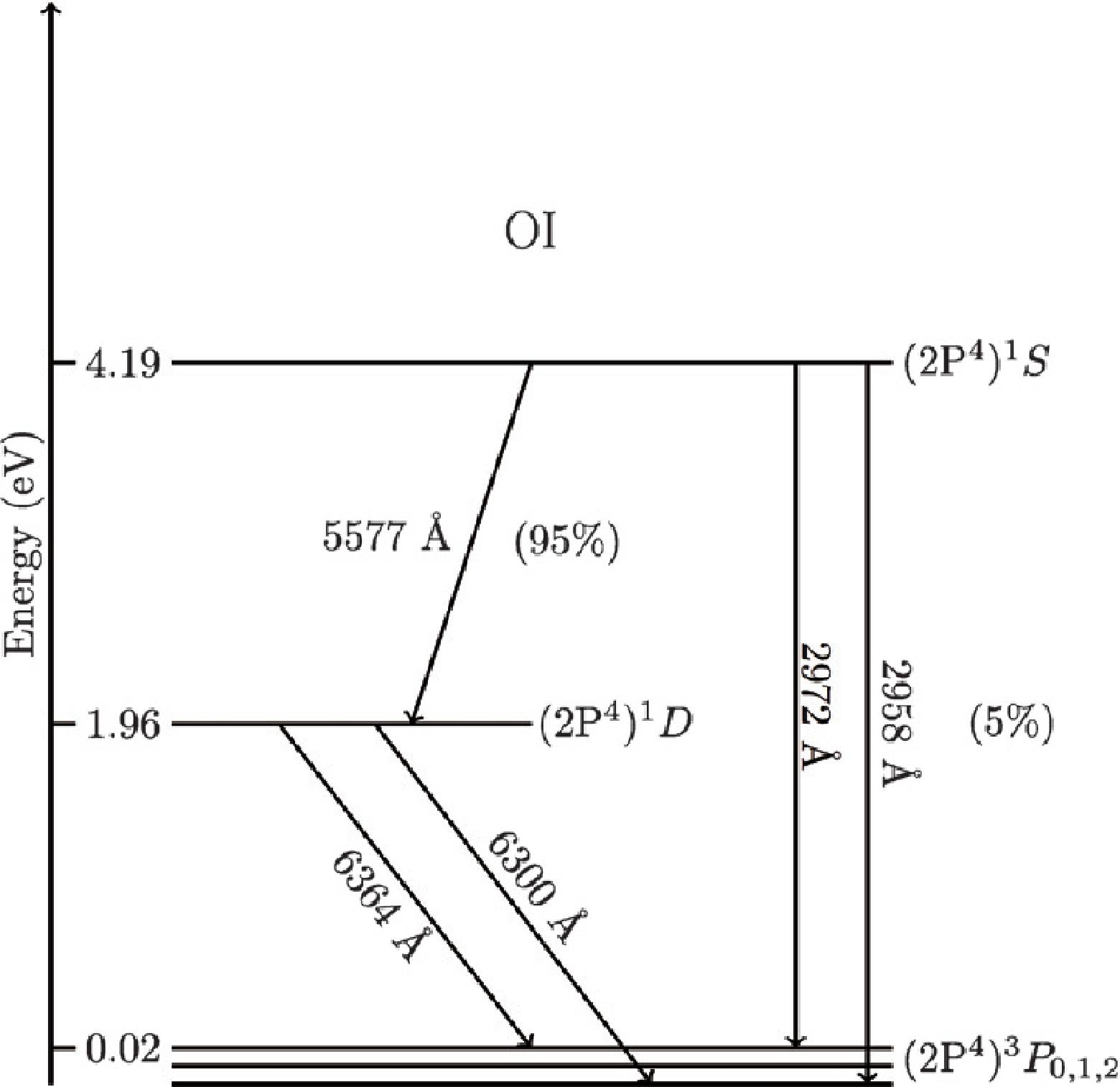}
\caption{
\label{energy}
}
\end{center}
\end{figure}

Fig.~\ref{energy}: Energy level diagram for \ion{O}{1}.  Note that all oxygen atoms that radiatively decay through the 5577~\AA~line then subsequently decay through either the 6300~\AA~or 6364~\AA~line.  Image credit~\cite{BhardwajRaghuram2012}.

\begin{figure}[p!]
\begin{center}
\includegraphics[height=\linewidth, angle=90]{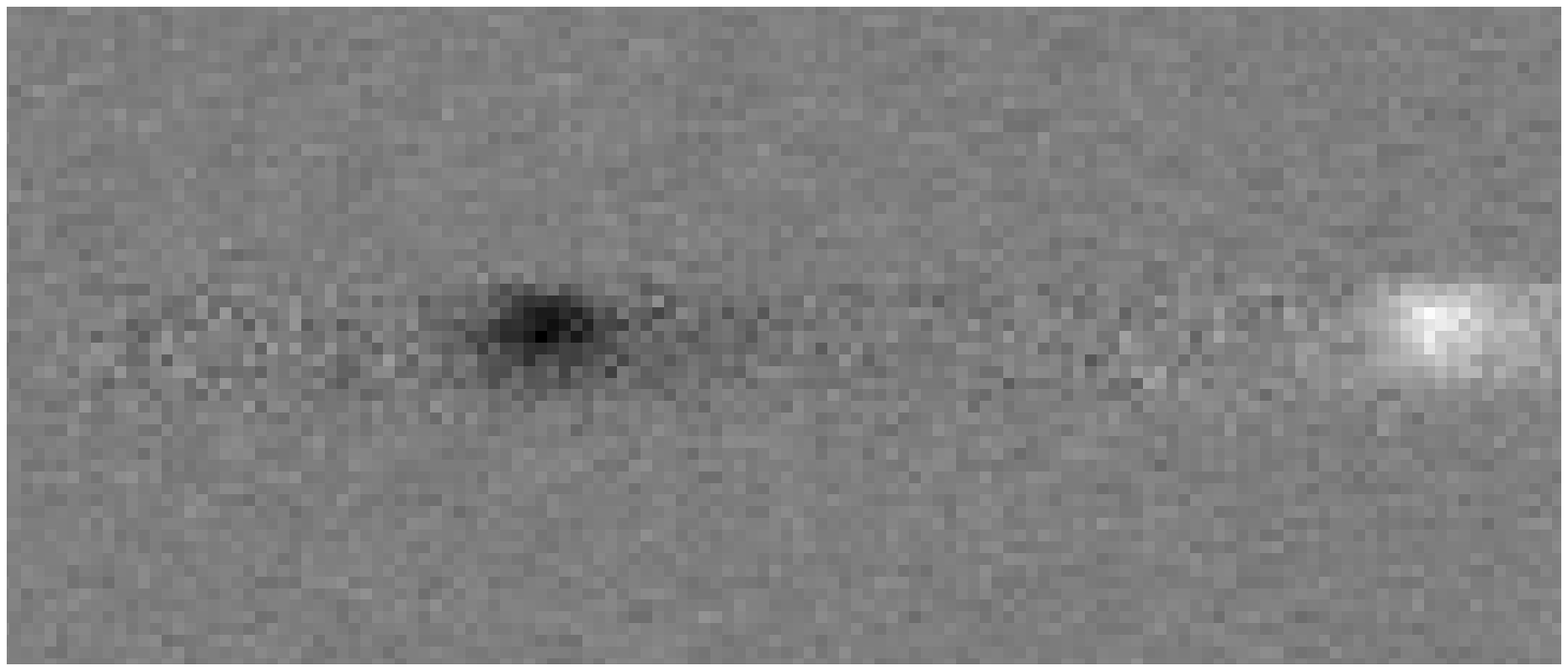}
\caption{
\label{CSHELLslit}
}
\end{center}
\end{figure}

Fig.~\ref{CSHELLslit}: A-B image showing the slit position on the comet for our UT October 10, 2011 observations.  The A-beam is the bright spot near the right edge and the B-beam is the dark spot to the left.  The location of the beams shifts along the slit (to the right in the above figure) when in imaging mode as compared to spectral mode, so the beams in the spectra are displaced 20-30 pixels to the left along the slit compared to the above images.

\begin{figure}[p!]
\begin{center}
\includegraphics[width=\linewidth]{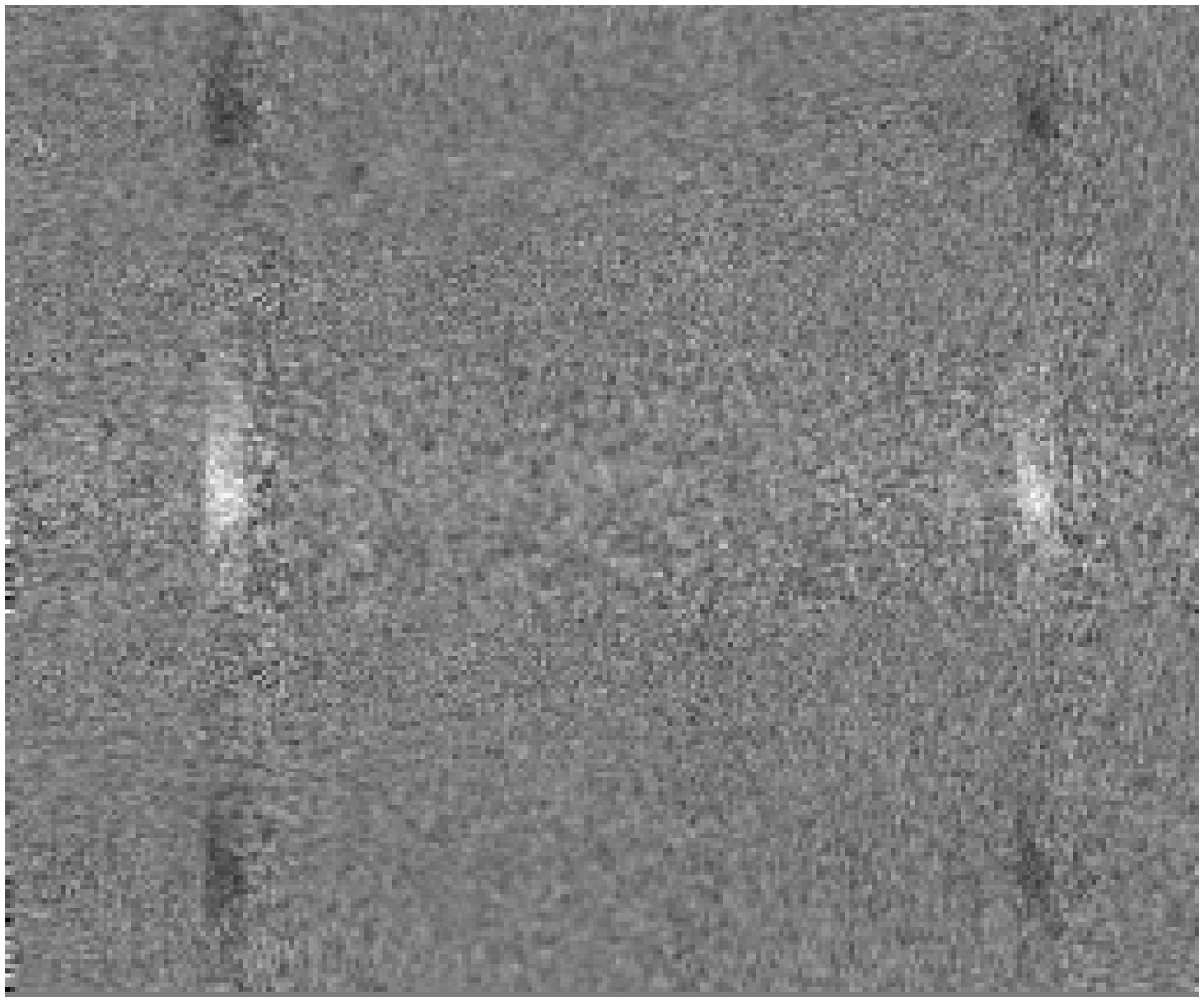}
\caption{
\label{CSHELLspec}
}
\end{center}
\end{figure}

Fig.~\ref{CSHELLspec}: A stacked and A-B subtracted spectrum in the CO grating setting of comet Garradd taken on UT January 26, 2012 with CSHELL (32 minutes on source).  The CO emission is clearly seen as the two bright spots, H$_2$O emission is much fainter, and is just to the left of the right CO line.  The dark spots correspond to the position of CO emission in the B beam used for subtraction of the background. 

\begin{figure}[p!]
\begin{center}
\includegraphics[width=\linewidth]{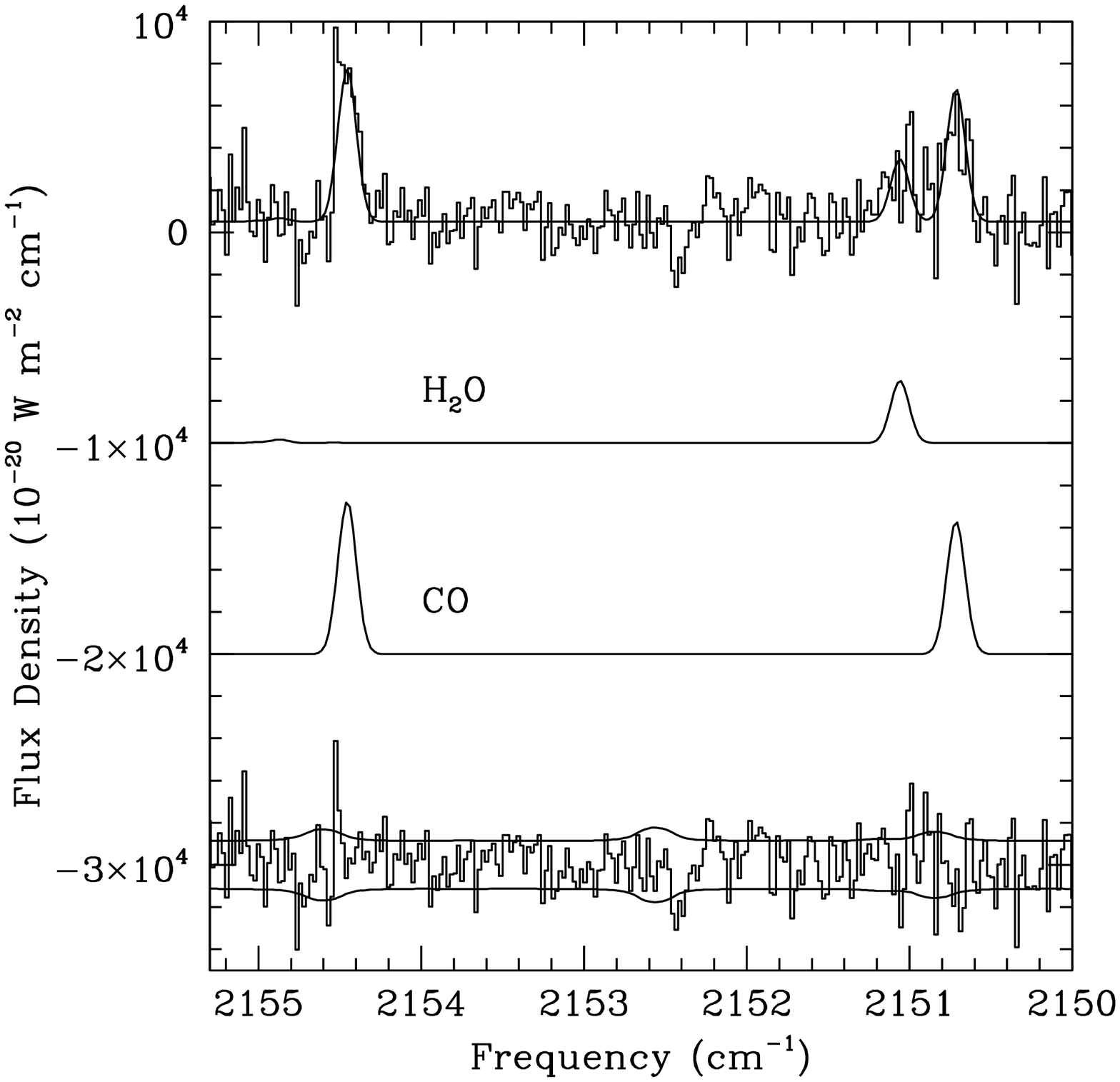}
\caption{
\label{COfit}
}
\end{center}
\end{figure}

Fig.~\ref{COfit}: Spectral fit to the CO setting for Garradd on October 10, 2011.  The data are plotted as a histogram at the top with the fit overplotted.  Offset below the data are the individual fits to the H$_2$O and CO emission lines.  At the bottom we show the residuals and the 1-sigma error envelope.

\begin{figure}[p!]
\begin{center}
\includegraphics[width=\linewidth]{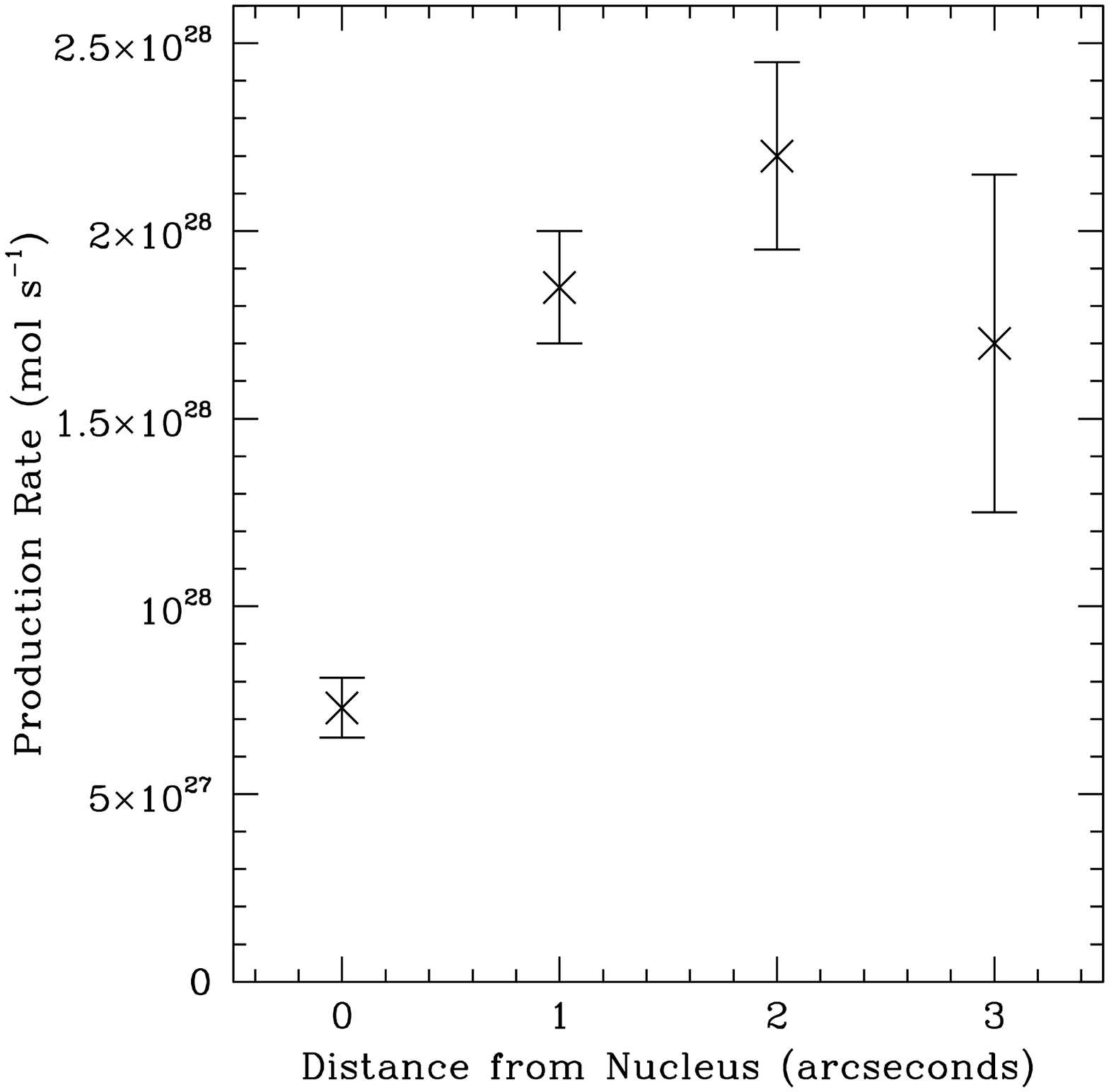}
\caption{
\label{QcurveIR}
}
\end{center}
\end{figure}

Fig.~\ref{QcurveIR}: Q-curve for Garradd on UT March 21, 2012. Inferred production rates increase as one moves the extraction aperture away from the nucleus and converges toward a global value.  The slit length is 30\arcsec.  However, the SNR falls off rapidly, meaning only production rates derived from the inner few arcseconds are meaningful.

\begin{figure}[h!]
\begin{center}
\includegraphics[width=\linewidth]{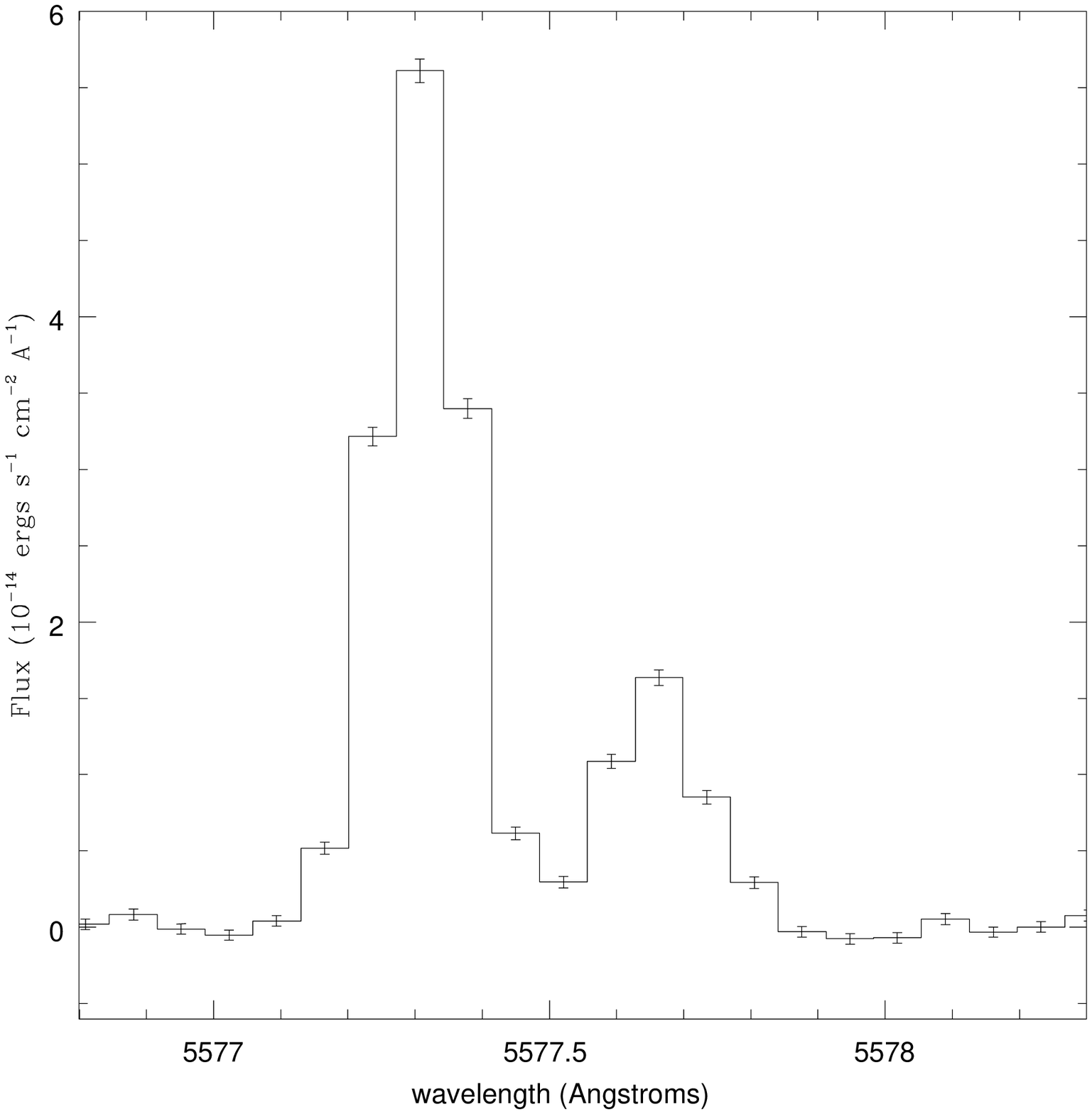}
\caption{
\label{OIspec}
\label{lastfig}
}
\end{center}
\end{figure}

Fig.~\ref{OIspec}: [\ion{O}{1}]5577 emission on UT September 21, 2011.  The cometary line is the weaker line to the right of the telluric line and is clearly resolved from the telluric counterpart.

\begin{figure}[p!]
\begin{center}
\includegraphics[width=\linewidth]{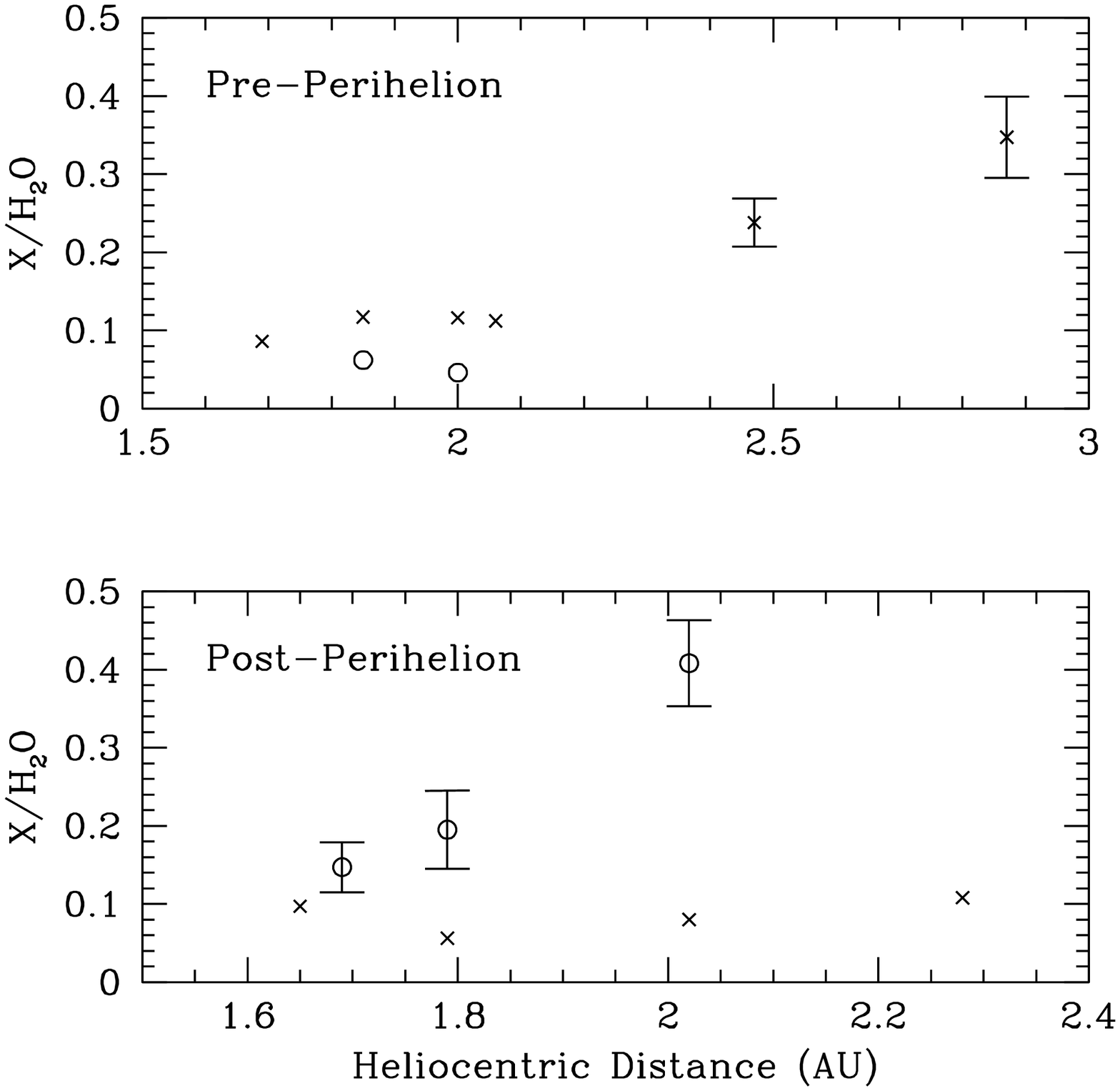}
\caption{
\label{COmixrat}
\label{lastfig}
}
\end{center}
\end{figure}

Fig.~\ref{COmixrat}: CO/H$_2$O (circles) from our CSHELL observations and CO$_2$/H$_2$O ($\times$) as derived from the \ion{O}{1} observations for Garradd as a function of heliocentric distance.  Here, and in subsequent figures, for the points without error bars, the uncertainty is similar to or smaller than the symbol.  CO and CO$_2$ exhibit different mixing ratios compared to H$_2$O throughout the apparition.

\begin{figure}[h!]
\begin{center}
\includegraphics[width=\linewidth]{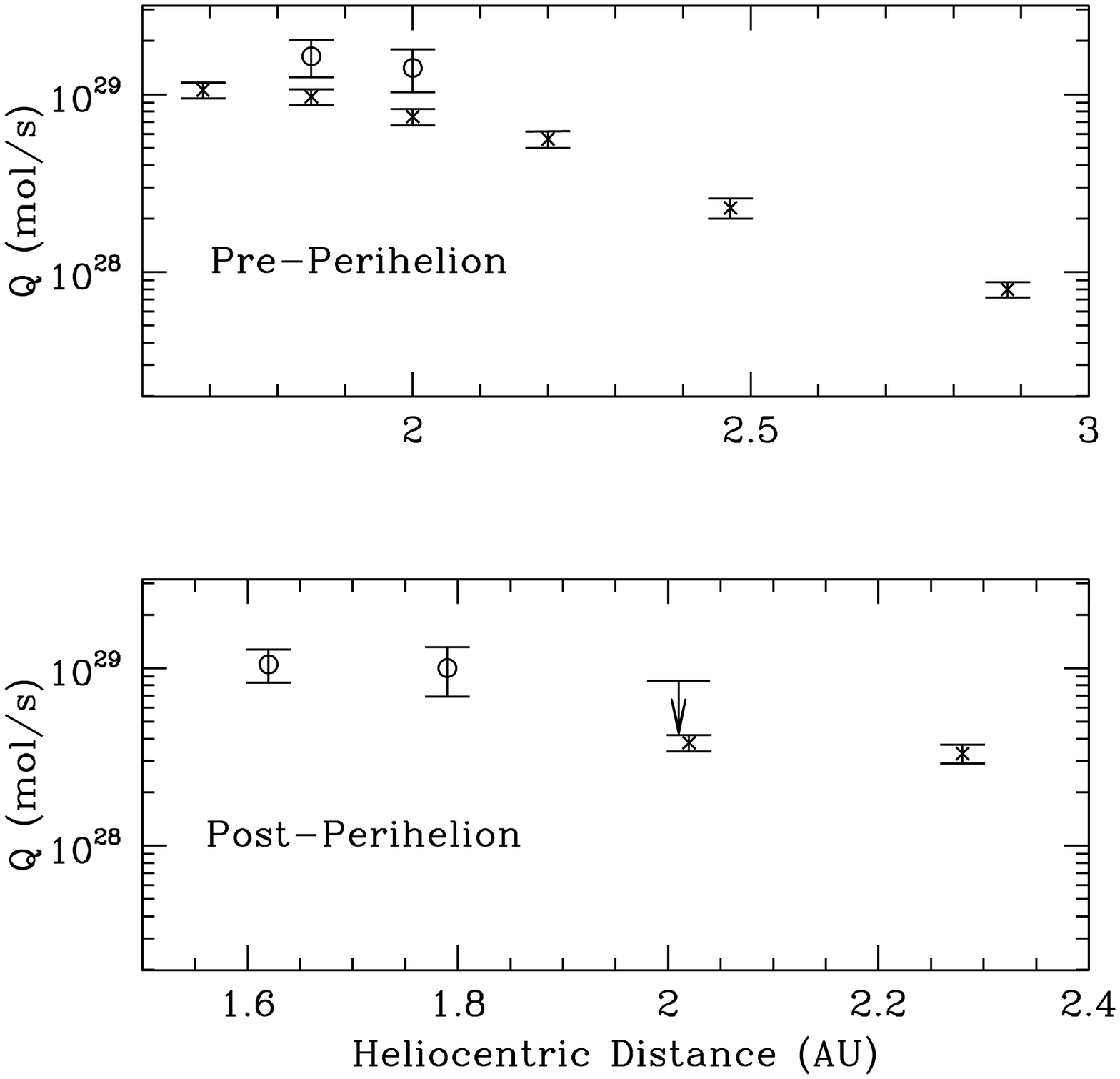}
\caption{
\label{QH2O_opt}
\label{lastfig}
}
\end{center}
\end{figure}

Fig.~\ref{QH2O_opt}: H$_2$O production rates as derived from the \ion{O}{1} observations ($\times$) and from the CSHELL observations (circles) as a function of heliocentric distance.  The upper limit arrow indicates our 3-sigma upper limit on H$_2$O production for our March CSHELL observations.  There is a tendency for the H$_2$O production rates derived from the CSHELL observations to be higher than those derived from the \ion{O}{1} observations pre-perihelion; it is unclear whether this is the case post-perihelion.

\begin{figure}[h!]
\begin{center}
\includegraphics[width=\linewidth]{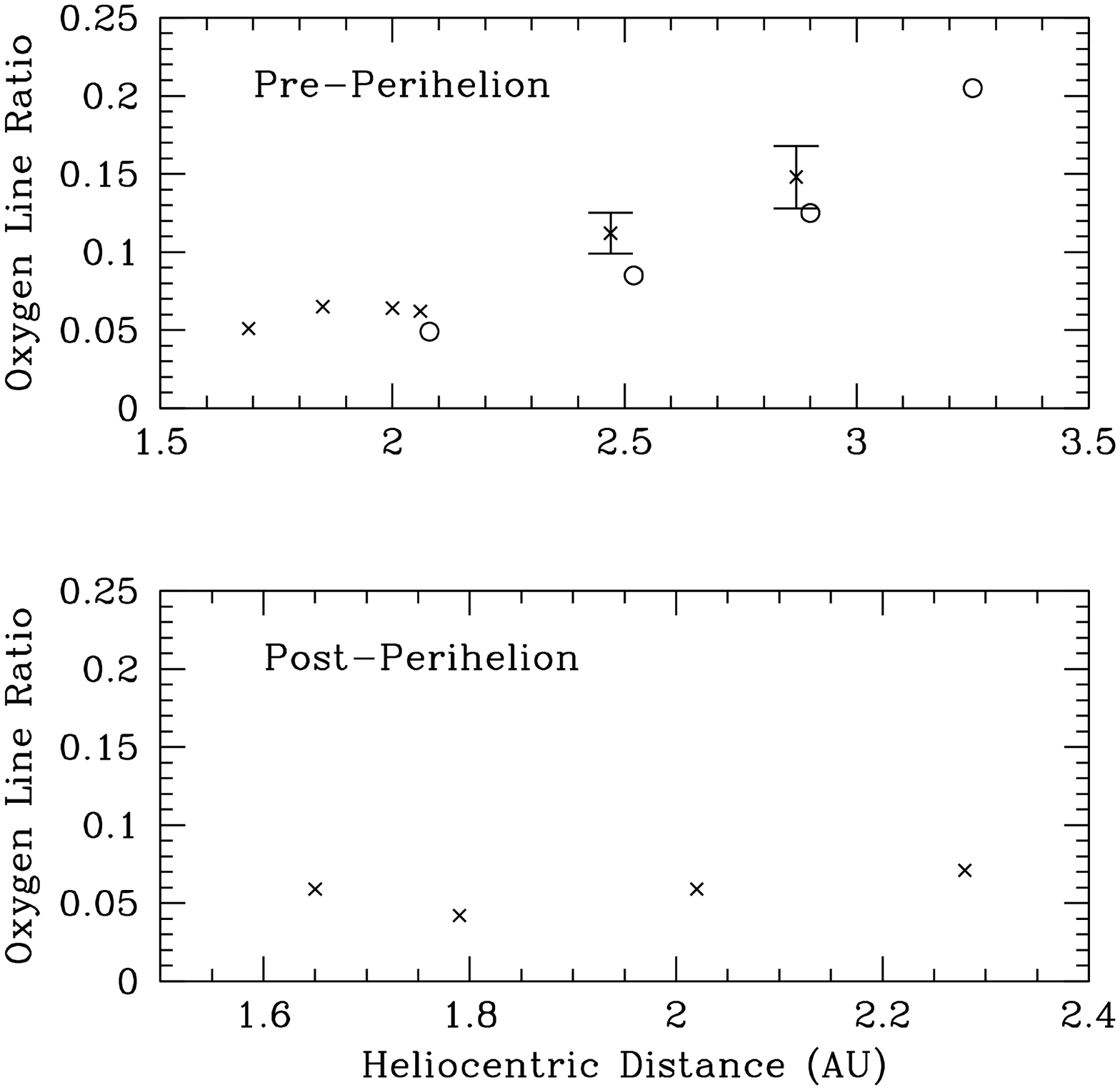}
\caption{
\label{OIratiofig}
\label{lastfig}
}
\end{center}
\end{figure}

Fig.~\ref{OIratiofig}: \ion{O}{1} line ratios for Garradd (with collisional quenching) as a function of heliocentric distance.  Our points are depicted by $\times$'s, while measurements from~\cite{Decock2013} are shown as circles.  All values have been corrected for collisional quenching.

\begin{figure}[h!]
\begin{center}
\includegraphics[width=\linewidth]{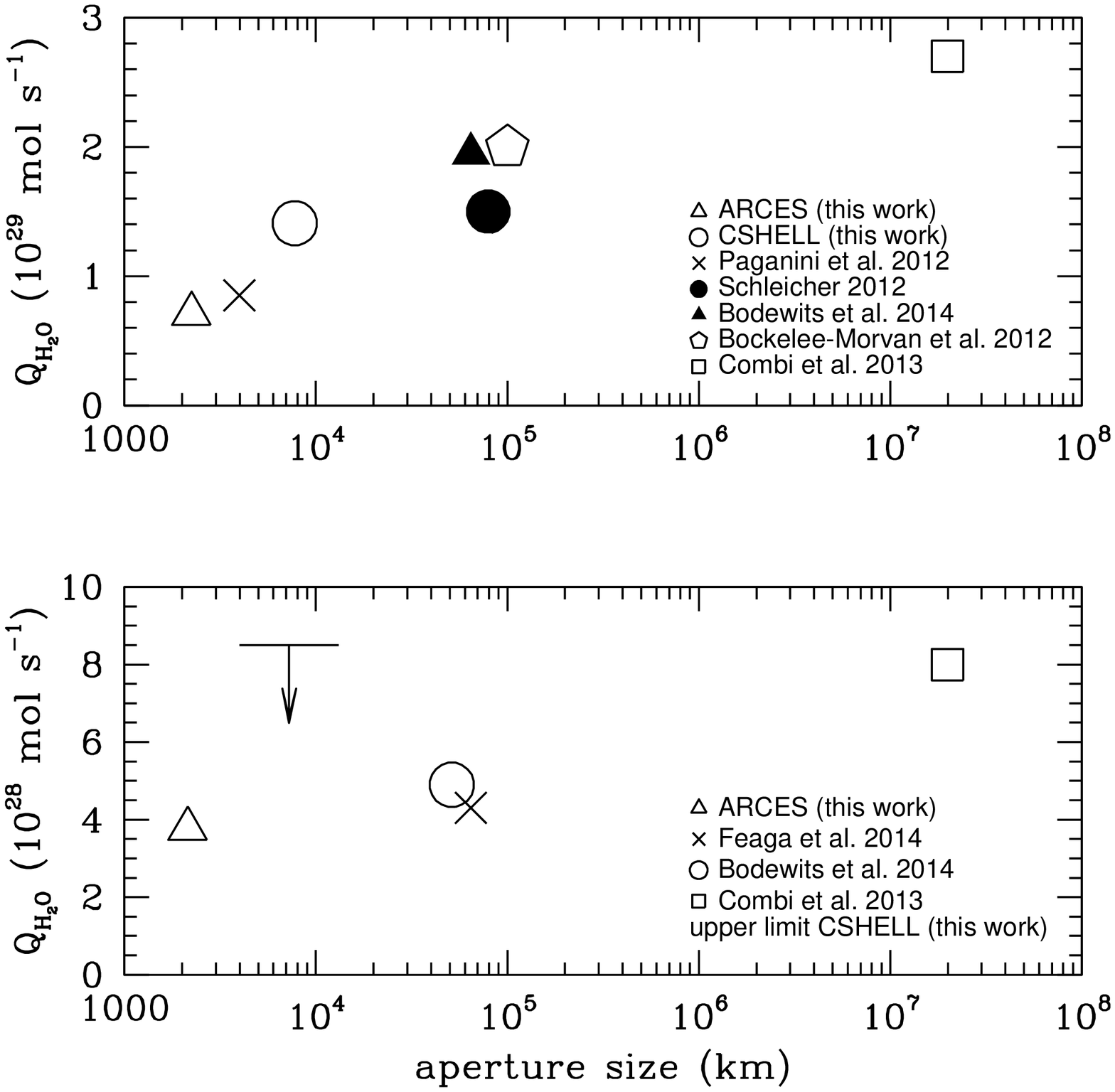}
\caption{
\label{aperture}
\label{lastfig}
}
\end{center}
\end{figure}

Fig~\ref{aperture}: Upper Panel: H$_2$O production rate as a function of aperture size for Garradd at a heliocentric distance of 2 AU pre-perihelion.  Our ARCES observation is denoted by a triangle and the CSHELL observation by a circle.  The other values in are taken from~\cite{Combi2013} and have the following sources: ~\cite{Bodewits2014} (filled triangle),~\cite{Combi2013} (square),~\cite{BockeleeMorvan2012} (pentagon), Schleicher 2012 (private communication) (filled circle),~\cite{Paganini2012} ($\times$).  The uncertainties are comparable to the size of the points.  There is a trend for larger aperture size observations to measure higher production rates, which is evidence that a significant fraction of the H$_2$O production is coming from icy grains in the coma.  Lower Panel: H$_2$O production rate as a function of aperture size for Garradd at a heliocentric distance of 2 AU post-perihelion.  Our ARCES observation is shown as a triangle and the CSHELL observation as an upper limit arrow.  The other values are taken from~\cite{Bodewits2014} (circle),~\cite{Combi2013} (square), and~\cite{Feaga2014} ($\times$).  The aperture effect observed pre-perihelion still seems to be present, but to a lesser degree. 

\clearpage

\end{document}